\documentclass[a4paper,11pt]{article}
\usepackage{jheppub} 
\usepackage{lineno}
\usepackage{bbm}
\usepackage{tikz}
\usetikzlibrary{calc}
\usepackage{graphicx,subcaption}
\usepackage{mathtools}
\usepackage{bm}
\usepackage{enumitem}
\usepackage{verbatim}
\usepackage{amssymb}
\usepackage{slashed}
\usepackage{float}
\usepackage[scr = dutchcal]{mathalpha}
\usepackage{lipsum}
\usepackage{multirow}
\usepackage{hyperref}
\usepackage[normalem]{ulem}
\usepackage{orcidlink}

\usepackage{circuitikz}

\newcommand{\blank}[1]{}

\renewcommand{\Re}{\operatorname{Re}}
\renewcommand{\Im}{\operatorname{Im}}

\DeclareMathOperator{\Tr}{Tr}

\DeclareMathAlphabet{\mathbfsf}{OT1}{cmss}{bx}{n}

\newcommand{\CA}{\mathcal{A}}
\newcommand{\CB}{\mathcal{B}}
\newcommand{\CC}{\mathcal{C}}
\newcommand{\CD}{\mathcal{D}}
\newcommand{\CE}{\mathcal{E}}
\newcommand{\CF}{\mathcal{F}}

\newcommand{\CH}{\mathcal{H}}
\newcommand{\CI}{\mathcal I}

\newcommand{\CO}{\mathcal{O}}

\newcommand{\CR}{\mathcal{R}}

\newcommand{\CW}{\mathcal{W}}
\newcommand{\CX}{\mathcal{X}}
\newcommand{\CY}{\mathcal{Y}}

\newcommand{\SCB}{\mathscr{B}}

\newcommand{\SCH}{\mathscr{H}}

\newcommand{\SCQ}{\mathscr Q}

\newcommand{\SCW}{\mathscr{W}}

\newcommand{\bb}[1]{\mathbb{#1}}

\renewcommand{\Im}{\text{Im}}
\renewcommand{\Re}{\text{Re}}

\renewcommand{\a}{\alpha}

\renewcommand{\t}{\tau}
\newcommand{\g}{\gamma}
\newcommand\G{\Gamma}
\newcommand\om{\omega}

\newcommand{\del}{\partial}
\newcommand{\that}{\hat{\tau}}

\renewcommand{\tilde}{\widetilde}

\newcommand{\Q}{\mathbf{Q}}
\newcommand{\I}{\mathbf{I}}
\newcommand{\J}{\mathbf{J}}

\newcommand{\slz}{\text{SL}(2,\mathbb{Z})}

\newcommand{\vev}[1]{\langle {#1} \rangle } 
\newcommand{\ket}[1]{|#1\rangle}

\newcommand{\h}{\mathbb{H}}
\newcommand{\be}{\begin{equation}}
\newcommand{\ee}{\end{equation}}

\arxivnumber{} 

\title{\boldmath Modular Properties of Symplectic Fermion Generalised Gibbs Ensembles}

\author{Faisal Karimi \orcidlink{0009-0003-5591-9911}}
\author{and G\'erard M.T. Watts \orcidlink{0000-0002-9066-2838}}
\affiliation{Department of Mathematics,\\
King's College London,\\
Strand, London, WC2R 2LS, United Kingdom\\}

\emailAdd{faisal.karimi@kcl.ac.uk}
\emailAdd{gerard.watts@kcl.ac.uk}

\abstract{
The symplectic fermion 
is a much-studied non-unitary conformal field theory with $c=-2$ and is known to contain an infinite family of mutually commuting conserved charges.
We derive expressions for the conserved charges on the cylinder and use these to 
construct Generalised Gibbs Ensembles (GGEs) in the particular case of the $\mathcal{W}(1,2)$ triplet model. We derive exact expressions for the modular \(S\)-transforms in each sector of the symplectic fermion (and so of the whole GGE) and further extract the expressions in the asymptotic regime where the chemical potentials go to zero.
Subsets of the conserved charges are known to reproduce the KdV and Boussinesq hierarchies. For the case in which the charge is identified with the zero mode \(W_0\) of the \(W_3\) algebra, we obtain asymptotic behaviour in precise agreement with the conjecture proposed in our companion paper \cite{Downing:2025huv}; for the KdV subset we obtain results which exactly mirror the case for a single free fermion. Finally we identify the GGE with a translation invariant and purely transmitting defect for the symplectic fermion fields, and make some comments on the relation to other $W_n$ algebras.
}

\begin{document}
\maketitle
\flushbottom
\section{Introduction}
All two-dimensional conformal field theories (2D CFTs) possess infinite sets of conserved currents and corresponding conserved charges\footnote{We exclude the trivial CFT with central charge $c=0$} as every holomorphic field is a conserved current and leads to a conserved charge. More surprisingly, it is also always possible to find infinite sets of charges which mutually commute, since the Virasoro algebra gives rise to the well-known KdV hierarchy of charges (and two further distinct hierarchies). If a theory has an extended symmetry, there can be further hierarchies - the $W_3$ algebra gives rise to the Boussinesq hierarchy (as well as other hierarchies) \cite{Kupershmidt:1989bf, Bazhanov:2001xm,Ashok:2024zmw}, and in general there are hierarchies associated with every affine Lie algebra \cite{Feigin:1991qm}.

In this work we focus on the symplectic fermion with \hbox{$c=-2$} and in particular on the $\CW(1,2)$ triplet model that can be constructed using it.
This model has been studied for a long time, and an infinite set of fields of all integer weights greater than 1 have been constructed \cite{Pope:1989ew,Shen:1992dd} (called the linear $W_\infty$ algebra) which are bilinear in the symplectic fermion field and whose zero modes commute. Subsets of these charges have further been identified with the KdV and Boussinesq hierarchies \cite{DiFrancesco:1991xm}.

Given a set of commuting conserved quantities, it is natural to replace the standard partition function, representing the standard Gibbs ensemble for a system at finite temperature, by a Generalised Gibbs ensemble (GGE) where chemical potentials are introduced for each of the charges.

The standard partition function of a CFT is defined as a trace over the full Hilbert space $\cal H$. 
If the system is formulated on a circle of 
length $R$, with the Hamiltonian $H$ and momentum generator $P$ given by the integrals around the cylinder of the conserved energy and momentum densities, 
then the partition function is given as
\begin{align}
    Z = \Tr_{\CH}(
    e^{-\beta H + i \gamma P} 
    ) \;,
\end{align}
where $\beta=1/T$ and $\gamma$ is to accommodate systems with non-zero total momentum.
To calculate this expression, one must map the energy and momentum densities to the plane with the result \begin{align}
    H = \frac{2\pi}R(L_0 + \bar L_0 - \frac{c}{12})
    \;,\;\;\;\;
    P = \frac{2\pi}{R}(L_0 - \bar L_0)\;.
\end{align}
where $L_0$ and $\bar L_0$ are elements of the Virasoro algebra. The partition function can then be interpreted as the partition function on a torus with complex structure parametrised by $\tau$,
\begin{align}
\label{eq:Z}
Z = 
    \Tr_{\cal H}(q^{L_0 - c/24} \bar q^{\bar L_0 - c/24})\;.
\end{align}
where $q=\exp(2\pi i \tau)$ and 
$\tau = (\gamma + i\beta)/R$.

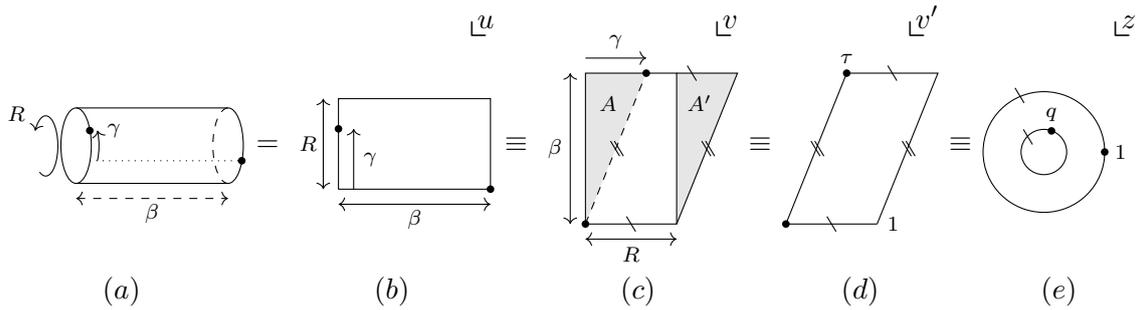
\begin{figure}[htb]
\[
\begin{array}{cccccccccc}
&&
\hfill\mbox{\Large$\llcorner$}\raisebox{1mm}{\kern -1.4mm{$u$}}
&&
\hfill\mbox{\Large$\llcorner$}\raisebox{1mm}{\kern -1.4mm{$v$}}
&&
\hfill\mbox{\Large$\llcorner$}\raisebox{1mm}{\kern -1.4mm{$v'$}}
&&
\hfill\mbox{\Large$\llcorner$}\raisebox{1mm}{\kern -1.4mm{$z$}}
\\[-3mm]
\begin{tikzpicture}[baseline=2em]
\begin{scope}[yshift=1.4em]
\draw (0,0.2) -- (2.,0.2);
\draw (0,1.2) -- (2.,1.2);
\draw (0,0.7) ellipse (0.2 and 0.5);
\draw (2.,0.2) arc (-90:90:0.2 and 0.5);
\draw[dashed] (2.,1.2) arc (90:270:0.2 and 0.5);
\draw[->] (-0.48,0.35) arc (-120:150:0.16 and 0.4) node [above left] {$\scriptstyle R$};
\draw[<->,dashed] (0,0) -- (2,0) ;
\node at (1,-0.2) {$\scriptstyle \beta$} ;
\fill[black] (2.18,0.5) circle (0.05);
\fill[black] (0.18,0.9) circle (0.05);
\draw[dotted] (0.18,0.5) -- (2.18,0.5);
\draw[->] (0.28,0.5) arc (-30:30:0.16 and 0.4) node [right] {$\scriptstyle \gamma$};
\end{scope}
\end{tikzpicture}
&
\raisebox{4mm}{\hbox{$=$}}
&
\begin{tikzpicture}[baseline=2em]
\begin{scope}[yshift=1.2em]
\draw (0,0.2) -- (2.,0.2) -- (2,1.4) -- (0,1.4) -- (0,0.2);
\draw[<->] (0,0) -- (2,0) ;
\draw[<->] (-0.2,0.2) -- (-0.2,1.4);
\node at (1,-0.2) {$\scriptstyle \beta$} ;
\node at (-0.4,0.8) {$\scriptstyle R$} ;
\fill[black] (2.,0.2) circle (0.05);
\fill[black] (0.,1.0) circle (0.05);
\draw[->] (0.2,0.2) -- (0.2,0.6) node [right] {$\scriptstyle \gamma$}  -- (0.2,1.0);
\end{scope}
\end{tikzpicture}
&
\raisebox{4mm}{\hbox{$\equiv$}}
&
\begin{tikzpicture}[baseline=2em]
\fill[gray!20] (0,0.2) -- (0.8,2.2) -- (0,2.2) -- (0,0.2);
\node at (0.3,1.8) {$\scriptstyle A$};
\fill[gray!20] (1.2,0.2) -- (2.0,2.2) -- (1.2,2.2) -- (1.2,0.2);
\node at (1.5,1.8) {$\scriptstyle A'$};
\draw (0,0.2) -- (1.2,0.2) -- (1.2,2.2) -- (0,2.2) -- (0,0.2);
\draw[<->] (0,0) -- (1.2,0) ;
\draw[<->] (-0.2,0.2) -- (-0.2,2.2);
\node at (0.6,-0.2) {$\scriptstyle R$} ;
\node at (-0.4,1.2) {$\scriptstyle \beta$} ;
\fill[black] (0,0.2) circle (0.05);
\fill[black] (0.8,2.2) circle (0.05);
\draw[->] (0,2.4) -- (0.4,2.4)  node [above] {$\scriptstyle \gamma$} -- (0.8,2.4);
\draw[dashed] (0,0.2) -- (0.8,2.2);
\draw (1.2,0.2) -- (2.0,2.2) -- (1.2,2.2);
\draw (0.54,0.3)--(0.66,0.1);
\draw (1.34,2.3)--(1.46,2.1);
\draw (0.33,1.3)--(0.45,1.1);
\draw (0.38,1.3)--(0.5,1.1);
\draw (1.53,1.3)--(1.65,1.1);
\draw (1.58,1.3)--(1.70,1.1);
\end{tikzpicture}
&
\raisebox{4mm}{\hbox{$\equiv$}}
&
\begin{tikzpicture}[baseline=2em]
\draw (0,0.2) -- (1.2,0.2)  node [right] {$\scriptstyle 1$};
\fill[black] (0,0.2) circle (0.05);
\fill[black] (0.8,2.2) circle (0.05) node [above] {$\scriptstyle\tau$};
\draw (0,0.2) -- (0.8,2.2);
\draw (1.2,0.2) -- (2.0,2.2) -- (0.8,2.2);
\draw (0.54,0.3)--(0.66,0.1);
\draw (1.34,2.3)--(1.46,2.1);
\draw (0.33,1.3)--(0.45,1.1);
\draw (0.38,1.3)--(0.5,1.1);
\draw (1.53,1.3)--(1.65,1.1);
\draw (1.58,1.3)--(1.70,1.1);
\end{tikzpicture}
&
\raisebox{4mm}{\hbox{$\equiv$}}
&
\begin{tikzpicture}[baseline=2em]
\begin{scope}[yshift=3em]
\draw (0,0) circle (0.8);
\draw (0,0) circle (0.3);
\fill[black] (0.8,0) circle (0.05) node [right] {$\scriptstyle 1$};;
\fill[black] (0.1,0.28) circle (0.05) node [above] {$\scriptstyle q$};;
\draw (-0.38,0.85)--(-0.26,0.65);
\draw (-0.27,0.3)--(-0.15,0.1);
\end{scope}
\end{tikzpicture}
\\
(a) && (b) && (c) && (d) && (e) \end{array}
\]
\caption{(a) The cylinder of length $\beta$, circumference $R$ with a
  translation through $\gamma$ before the identification of the ends
can be realised as a quotient of the plane (b) and
 (c), and is equivalent (after rescaling) to a
torus (d) with modular parameter $\tau$ and an annulus (e) with coordinates
$u,v,v',z$ respectively. The solid points are identified in each picture. This figure has been taken and modified from \cite{Downing:2021mfw}.
}
\label{fig: Figure from 2021}
\end{figure}

We show this in figure \ref{fig: Figure from 2021}. The cylinder is length $\beta$ and this can be seen as a quotient of the plane in coordinates $u$. Rotating to $v=-iu +R$ and rescaling to $v'=v/R$, we end up with a torus with the modular parameter $\tau = (\gamma + i \beta)/R$. This can then be mapped to an annulus by $z=\exp{(2\pi i v')} = \exp(2\pi u/R)$.

Given a set of conserved currents $J_i$ with corresponding charges $Q_i$, a simple generalisation of \eqref{eq:Z} is
\begin{align}
    \Tr_{\cal H}\left(e^{\sum_i \mu_i Q_i}\right)\;,
    \label{eq:genz}
\end{align}
where we have incorporated the energy and momentum operators as 
$Q_1 = L_0=c/24$ and $Q_{-1}=\bar L_0-c/24$ and
the function
\eqref{eq:genz} is well-defined for suitable conditions on the $\mu_i$.

If the charges commute, then we can interpret \eqref{eq:genz} as a GGE and the constants $\mu_i$ as chemical potentials.

GGEs are of interest because they play an important role in understanding the thermalisation properties of many body systems with additional conserved quantities, see \cite{DAlessio:2015qtq}. They are also related to understanding higher-spin black holes \cite{Gutperle:2011kf,  Gaberdiel:2012yb,Dymarsky:2020tjh}.

In 2D CFTs, the partition function is a modular invariant. This raises the natural question: does the GGE have any ``nice'' modular properties? 
The question of the modular properties of thermal correlation functions and GGEs has been of interest for a long time. See \cite{Gutperle:2011kf,Gaberdiel:2012yb,Dymarsky:2020tjh} for the context of higher spin black holes, \cite{Iles:2014gra} for an earlier look at insertions of $W_3$ charges, \cite{Maloney:2018hdg} for KdV thermal correlation functions, and \cite{Dymarsky:2018lhf, Dymarsky:2018iwx, Dymarsky:2022dhi, Brehm:2019fyy} for the large central charge limit of KdV GGEs. 

This problem has been studied previously at $c=\frac{1}{2}$ \cite{Downing:2021mfw,Downing:2023lnp,Downing:2023lop}, where exact expressions for the modular transform of GGEs with KdV charges were obtained and interpreted in terms of defect insertions, and it was shown that 
the modular transform has the asymptotic expansion as a GGE where the charges are in the same hierarchy -- but this is not the general result. 
The modular properties of KdV GGEs were investigated more broadly for KdV charges using the Thermodynamic Bethe Ansatz in \cite{Downing:2024nfb} where it was shown that, in general, the modular transform of a GGE is not another GGE. The asymptotic modular transform takes the form \eqref{eq:genz} but for charges $\{Q_i\}$ which do not all mutually commute.

This is also true for a Boussinesq GGE - in our companion work \cite{Downing:2025huv}, we 
considered the particular GGE
\begin{align}
    \Tr_{\cal H}(q^{L_0 - c/24} \bar q^{\bar L_0 - c/24}\, e^{\alpha W_0})\;,
\end{align}
 and proposed an expression for the modular transformation of this GGE in the regime $\alpha\to 0$, which was again of the form \eqref{eq:genz} but where the $Q_i$ do not, in general, all mutually commute (the proposed expression was conjectured in \cite{Downing:2025huv} and has now been proven \cite{proof}). 

In this paper, we 
consider the symplectic fermion which has $c=-2$. One can define various theories constructed from the symplectic fermion but we will restrict ourselves to the simplest modular invariant theory, the $\CW(1,2)$ triplet model \cite{Gaberdiel:2007jv}.

Earlier evidence from \cite{Downing:2024nfb} suggested that this may be a special case where the thermal correlators of KdV charges close under modular transformations, which is a pre-requisite for the asymptotic modular transform of the GGE to be itself a GGE of the same hierarchy. We confirm that this is the case at $c=-2$. Furthermore, we show that this is also true of a large set of charges $\{\Q_n\}$ associated to fields bilinear in the symplectic fermion, where $n\geq 1$ is the spin of the charges. 
The KdV charges are the subset of these of odd spin, $\{\Q_{2n-1}\}$. The full set are the integrals of the chiral (and anti-chiral) fields of the linear $W_\infty$ algebra identified in \cite{Shen:1992dd}.

Since the basis of currents in \cite{Shen:1992dd} differs from our normalisation, we 
re-derive the expressions for the currents in section \ref{subsec:basisoffields}.
We also need the charges, which are the result of mapping the integrals of the conserved currents on the cylinder to the plane. We do this in two independent ways in section \ref{subsec: Expressions for the charges}.
Since the resulting charges are bilinear in the fermion modes, their action on the Hilbert space is easy to calculate (although they are not all diagonal due to the complicated ground state structure of the triplet model) and it is easy to find the exact expression for the GGE.

Given the expression for the GGE, we are able to find and prove an exact expression for its modular transform, in the style of \cite{Downing:2021mfw,Downing:2024nfb}.
We show that, as with the free fermions at $c=1/2$, the modular transform of a GGE constructed from these charges has an asymptotic expansion as a GGE of the same set of charges and 
show that it can also be expressed as the partition function of the symplectic fermion system in the presence of a purely transmitting defect, as was the case for the free fermion.
We also show that in the particular case of the $W_0$ charge of the $W_3$ algebra, this agrees with the expressions in \cite{Downing:2025huv,proof}.

The paper is structured as follows:

Section \ref{sec: The Symplectic Fermion} reviews the Symplectic Fermion model, its representations, and its relation to the triplet model.
Although the triplet model and the symplectic fermion have been covered extensively in the literature, we think it is helpful to review those aspects we need, in particular the untwisted and half-twisted sectors, which are central to our modular analysis, define their (super)characters and give their modular properties.

In Section \ref{sec: Hierarchy of Bilinear Fields}, we construct a hierarchy of bilinear fields in the Symplectic Fermion theory. We provide two independent derivations of the explicit forms of their associated charges. In Section \ref{subsec: Realisation of Hierarchies} we establish the connection to known integrable hierarchies. In connection with this, it is demonstrated that the corresponding currents satisfy the pre-Lie algebra structure identified by Dijkgraaf \cite{Dijkgraaf:1996iy} which, while important for the link to our companion paper, is not immediately relevant to this work and so it has been relegated to appendix \ref{subsec: Pre-Lie}.

In Sections \ref{sec:The Symplectic Fermion GGE} and \ref{sec: Exact Transformation of Symplectic Fermion GGE} we define the Generalised Gibbs Ensemble (GGE) of the Symplectic Fermion by inserting charges from the bilinear hierarchy into the partition function and the exact modular $S$-transformation of the Symplectic Fermion GGE is derived in Section \ref{sec: Exact Transformation of Symplectic Fermion GGE}.

Section \ref{sec: Asymptotic Analysis of Symplectic Fermion GGE} provides analyses of the transformed GGE in the asymptotic regime where the chemical potential approaches 0. It is shown that the exact result is consistent in this regime with perturbative checks, and a connection is made with \cite{Downing:2025huv}. A conjecture from \cite{Downing:2024nfb} is resolved about which central charges allow for the thermal correlators of KdV charges to close under modular transformations.

In section \ref{sec:Wn} we relate the triplet model to the $W_3$ algebra and the $W_n$ algebras in general, and show that the results in this paper recover the asymptotic results conjectured in \cite{Downing:2025huv}.

Finally in section \ref{sec:defect} we show how the results we have found are consistent with the identification of the GGE as a translation invariant purely-transmitting defect for the symplectic fermion fields.

\section{The Symplectic Fermion and Representations}\label{sec: The Symplectic Fermion}
At central charge $c=-2$ there exists a rich free field theory known as the Symplectic Fermion \cite{Kausch:2000fu}. This theory provides a realisation of a wide class of logarithmic CFTs (see, e.g., \cite{Creutzig:2013hma}). In this work, we restrict attention to the sectors of the Symplectic Fermion with relatively ``simple'' modular properties, namely the untwisted and half-twisted sectors; these are directly related to the representations of the Triplet Model \cite{Gaberdiel:1998ps}.

The fundamental fields are fermionic currents $\chi^\alpha(z)$ ($\alpha = \pm$) with operator product expansions
\begin{equation}
    \chi^{\alpha}(z)\chi^{\beta}(w) \sim \frac{\epsilon^{\alpha\beta}}{(z-w)^2}
    ,
    \qquad 
    \epsilon^{+-} = -\epsilon^{-+} = 1,\; \epsilon^{++}=\epsilon^{--}=0\;,
    \label{eq:chiope}
\end{equation}
where we define the symbol ``$\sim$" to mean that we have dropped terms of $O(1)$. 

The fields $\chi^\pm$ have the mode expansions
\begin{align}
    \chi^\pm(z) = \sum_{m\in\mathbb Z \mp\lambda} \chi^\pm_m \,z^{-m-1}\;,
    \label{eq:chimode}
\end{align}
where we have labelled the sectors by $\lambda$ where $\lambda=0$ is the untwisted sector and $\lambda=1/2$ is the half-twisted sector.

The operator product expansions \eqref{eq:chiope} imply the mode algebra
\begin{equation}\label{eq:chicomm}
    \{\chi^{\alpha}_m,\chi^\beta_n\} = m\,\epsilon^{\alpha\beta}\,\delta_{m+n,0}.
\end{equation}
The fields $\chi^\pm(z)$ are Virasoro primary fields of weight $1$ with respect to modes $L_m$ of the stress tensor
\begin{equation}
    [L_m,\chi^\pm_n]=-n\chi^\pm_{m+n}
    \;,\;\;
    T(z) = \sum_{m\in\mathbb Z}L_m\,z^{-m-2}
    \;,\;\;
    T(z) = \big(\chi^-\chi^+\big)(z)
    \;.
\end{equation}
where the brackets denote conformal normal ordering of fields, see $\S$6.5 of \cite{DiFrancesco:1997nk}. In particular, in the untwisted sector we have
\begin{equation}
    L_m = \sum_{n=0}^\infty \chi^-_{m-n}\chi^+_n -
    \sum_{n=1}^\infty \chi^+_{-n}\chi^-_{m+n}
    \;,\;\;
    L_0 = \chi^-_0 \chi^+_0 + \ldots,
\end{equation}
We can also write an expression for $L_m$ valid in both sectors, depending on the choice of $r$ and $s$,
\begin{equation}
     L_{r+s} =    
     \tfrac{1}{2}r(r+1)\,\delta_{r+s,0}  
    +
     \sum_{j=0}^\infty 
    \big\{
         \chi^-_{s-1-j}\chi^{+}_{r+1+j}
         - 
         \chi^+_{r-j}\chi^{-}_{s+j}
    \big\}
\end{equation}
The Virasoro algebra has integer moding, so although $r$ and $s$ need not be integers, one requires that $r+s\in\bb Z$.
The existence of a GGE requires a well-defined partition function, in as much it is an extension of a partition function. As emphasised in \cite{Kausch:2000fu,Creutzig:2013hma}, the symplectic fermion and its many twisted representations can realise infinitely many modular invariants. However, the simplest case, the case which we will focus on, is the modular invariant triplet model. Thus, the representations (more specifically, characters of representations)
of the symplectic fermion algebra \eqref{eq:chicomm} of principal
interest to us are the irreducible vacuum module $\mathbb{L}_0$ (to be introduced below)
in which the modes in \eqref{eq:chimode} have integer labels, and the irreducible half-twisted module $\mathbb{L}_{1/2}$ in which the modes are half-integers. We will explain the connection now between the symplectic fermion and triplet model.

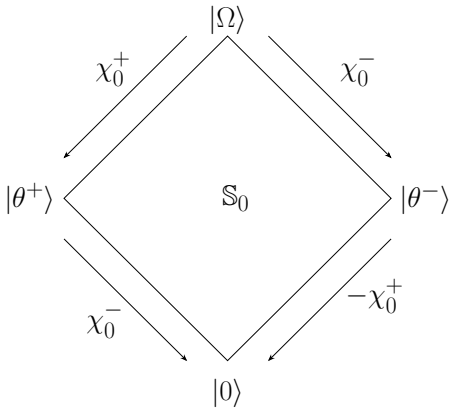
\begin{figure}[!ht]
\centering
\resizebox{0.4\textwidth}{!}{%
\begin{circuitikz}
\tikzstyle{every node}=[font=\Huge]

\draw  (16.25,11.75) -- (21.25,6.75) -- (26.25,11.75) -- (21.25,16.75) -- cycle;
\draw [->, >=Stealth] (20,16.75) -- (16.25,13);
\draw [->, >=Stealth] (22.5,16.75) -- (26.25,13);
\draw [->, >=Stealth] (26.25,10.5) -- (22.5,6.75);
\draw [->, >=Stealth] (16.25,10.5) -- (20,6.75);
\node [font=\Huge] at (21.25,17.25) {$|\Omega\rangle$};
\node [font=\Huge] at (15.2,11.75) {$|\theta^+\rangle$};
\node [font=\Huge] at (27.3,11.75) {$|\theta^-\rangle$};
\node [font=\Huge] at (21.25,5.75) {$|0\rangle$};
\node [font=\Huge] at (17.75,15.75) {$\chi_0^+$};
\node [font=\Huge] at (25.75,8.75) {$-\chi_0^+$};
\node [font=\Huge] at (25.25,15.75) {$\chi_0^-$};
\node [font=\Huge] at (17.5,8) {$\chi_0^-$};
\node [font=\Huge] at (21.5,11.75) {$\mathbb{S}_0$};
\end{circuitikz}
}
\caption{Ground state structure of the chiral indecomposable module $\mathbb{S}_0$ of the Symplectic Fermion.}
\label{fig: 1}
\end{figure}

First, the representations of the symplectic fermion. The irreducible vacuum module $\mathbb L_0$ is built on a primary state $\ket 0$
satisfying
\begin{align}
    \chi^\pm_m\ket 0 = 0\;,\;\; m\geq 0
\end{align}
and admits the Fock space construction \begin{equation}\label{eq: L_0}
    \bb{L}_0
= \text{span} \left\{ \chi^-_{-n_1}\dots\chi^-_{-n_\nu}\chi^+_{-m_1}\dots\chi^+_{-m_\mu}\ket{0} \,\,\textbf{;} \,
        n_i > n_{i{+}1},\\m_{i}>m_{i{+}1},
    \, n_i,m_i \in \bb{Z}{}^{\vphantom{|}}_{>0}\right\}\;.
\end{equation}
The half-twisted module $\mathbb L_{1/2}$
is built on a primary state $\ket{{-}\tfrac18}$ of conformal weight $h=-1/8$ satisfying
\begin{align}
    \chi^\pm_m \ket{{-}\tfrac 18} = 0\;,m>0
    \;.
\end{align}
and admits the Fock space construction
\begin{equation}\label{eq: L_1/2}
    \mathbb{L}_{1/2} 
    = \text{span} \left\{ \chi^-_{-n_1}\dots\chi^-_{-n_\nu}\chi^+_{-m_1}\dots\chi^+_{-m_\mu}\ket{{-}\tfrac 18} \,\,\textbf{;} \,
        n_i > n_{i{+}1},\\m_{i}>m_{i{+}1},
    \, n_i,m_i \in \bb{Z}^{\vphantom{!}}_{\geq 0}{+}\tfrac 12\right\}\;.
\end{equation}
In the untwisted sector, the existence of the zero modes $\chi^\pm_0$ allows one to consider larger, reducible but indecomposable representations, the largest of which we denote $\mathbb{S}_0$, containing $\mathbb{L}_0$ as a submodule. The ground state space of $\mathbb S_0$ is four-dimensional, consisting of the vacuum $\ket{0}$ together with its Jordan partner $\ket{\Omega}$ satisfying
\begin{equation}
    L_0 \ket{\Omega} = \ket{0},
\end{equation}
and the states $\ket{\theta^\pm}$ generated by the fermion zero-modes,
\begin{equation}
\ket{\theta^\pm} = \chi^\pm_0\ket \Omega
\;,\;\;
    \ket 0 = 
    \chi^-_0\chi^+_0\ket\Omega= \chi^-_0\ket{\theta^+}
    = - \chi^+_0\ket{\theta^-}.
\end{equation}
There are several further representations generated by subsets of the states $\{\ket{\theta^{\pm}}\}$, but these are reducible with respect to the symplectic fermion algebra; the representations $\bb{L}_0$ and $\bb{S}_0$ are special in that they are the irreducible representation and the maximal reducible representation respectively. Acting with
non-positive fermion modes on $\ket\Omega$ generates $\mathbb{S}_0$:
\begin{equation}\label{eq: S_0}
    \mathbb{S}_0 
    = \text{span} \left\{ \chi^-_{-n_1}\dots\chi^-_{-n_\nu}\chi^+_{-m_1}\dots\chi^+_{-m_\mu}\ket{\Omega} \,\,\textbf{;} \,
        n_i > n_{i{+}1},\\m_{i}>m_{i{+}1},
    \, n_i,m_i \in \bb{Z}{}^{\vphantom{|}}_{\geq0}\right\}\;.
\end{equation}
The structure of its ground states is illustrated in Fig.~\ref{fig: 1}.

\subsection{The Triplet Algebra}
\label{ssec:triplet}
The Triplet algebra, $\CW(1,2)$, is defined as the bosonic subalgebra of the symplectic fermion (the $\bb{Z}_2$ orbifold). It is generated by the stress tensor and three additional weight 3 fields
\begin{equation}
    W^{\pm}(z) = (\chi^\pm\del\chi^\pm)(z)\quad,\quad W(z) = \frac{1}{2}((\del\chi^- \chi^+) - (\chi^- \del \chi^+)).
\end{equation}
We can decompose the symplectic fermion representations into even and odd subspaces that form representations of the Triplet algebra.
\begin{equation}\label{eq: Even Odd Decompositions}
\begin{split}
    \bb{L}_0 = \SCW_0 \oplus \SCW_1 \quad&,\quad \bb{L}_{1/2} = \SCW_{-1/8} \oplus \SCW_{3/8}
\end{split}
\end{equation}
Here, the modules are labelled by the conformal dimensions of their ground states. The modules $\SCW_h$ is irreducible with respect to the Triplet algebra. The weight $0$ and weight $-1/8$ modules $\SCW_0$ and $\SCW_{-1/8}$ have one-dimensional ground state spaces, whereas $\SCW_1$ and $\SCW_{3/8}$ have two-dimensional ground state spaces.

The Hilbert space of the triplet model was constructed in \cite{Gaberdiel:1998ps,Gaberdiel:2007jv} and is of the form
\begin{equation}\label{eq: Triplet state space}
    \SCH_{\text{triplet}} = (\SCW_{-1/8}\otimes\overline{\SCW_{-1/8}})\oplus (\SCW_{3/8}\otimes\overline{\SCW_{3/8}})\oplus \SCB
    \;,
\end{equation}
where $\SCB$ is an appropriate quotient of the bosonic subspace of $\bb{S}_0 \otimes \overline{\bb{S}_0}$ and contains the irreducible triplet vacuum module $\SCW_0 \otimes \overline{\SCW_0}$. Rather than repeat the construction of $\SCB$, we will instead simply give a basis. We start from two bosonic states, which with a slight abuse of notation we call $\{\ket\Omega,\ket 0\}$, and two fermionic states $\{\ket{\psi^+},\ket{\psi^-}\}$ which are annihilated by all the positive symplectic fermion modes and satisfy 
\begin{align}
\begin{split}
  \ket{\psi^{\pm}} &= \chi_0^{\pm}\ket\Omega =  \overline{\chi}_0^{\pm}\ket\Omega
  \;,\;\;\;\; \\
\chi_0^{\a}\ket{\psi^\beta}&=\overline{\chi}_0^{\a}\ket{\psi^\beta} = \epsilon^{ \beta\alpha}\ket{0}
  \;,\;\;
  \epsilon^{+-} = -\epsilon^{-+} = 1\;,\;\;\epsilon^{++}=\epsilon^{--}=0\;,
  \\
    L_0\ket 0 &= L_0\ket{\psi^\pm}=\overline{L}_0\ket 0 = \overline{ L}_0\ket{\psi^\pm}=0\;, \;\;\;\;
    L_0 \ket{\Omega} = \overline{L}_0\ket{\Omega} = \ket{0}\;.
    \label{eq: Bulk Ground state relations}
\end{split}
\end{align}
Note that the states do not factorise into a tensor product of left and right moving modes acting on a single vacuum since $\ket{\psi^\alpha} = \chi^\alpha_0\ket 
\Omega=\bar\chi^\alpha_0\ket \Omega$. A basis of $\SCB$ is then given by
\begin{equation}
\SCB = V^e_{\ket{\Omega}}\oplus V^e_{\ket{0}}\oplus V^o_{\ket{\psi^{+}}}\oplus V^o_{\ket{\psi^-}}
\;,
\end{equation}
where $V^{e/o}_{\ket v}$ are given by
\begin{equation}
V^{e/o}_{\ket{v}} = \text{span}\{\chi^-_{-n_1}...\chi^-_{-n_{\nu}}\chi^+_{-m_1}...\chi^+_{-m_{\mu}}\overline{\chi}^-_{-\overline{n}_1}...\overline{\chi}^-_{-{\overline{n}}_{\sigma}}\overline{\chi}^+_{-\overline{m}_1}...\overline{\chi}^+_{-\overline{m}_{\rho}}\ket{v},\nu+\mu+\sigma+\rho = \text{even/odd} \},
\end{equation}
and $n_i>n_{i{+}1}$, $m_i>m_{i{+}1}$, $\bar n_i>\bar n_{i{+}1}$, $\bar m_i>\bar m_{i{+}1}$, and $n_i,m_i,\bar n_i,\bar m_i \in \bb{Z}_{>0}$. 

Let us turn to modular invariance - we give a brief introduction in appendix \ref{app: Modular Forms}. We will see that the modular invariant partition function of the triplet model state space \eqref{eq: Triplet state space} is expressible in terms of only characters and super-characters of the untwisted and half-twisted symplectic fermion modules $\bb{L}_0$ and $\bb{L}_{1/2}$ respectively.

We introduce the total fermion number operator, $(-1)^F$, which induces a $\bb{Z}_2$ grading on the relevant modules $\bb{L}_0$ and $\bb{L}_{1/2}$ and the quotiented space of $\bb{S}_0\otimes\overline{\bb{S}_0}$.
\begin{equation}
\begin{split}
    &\{\chi^\alpha_m,(-1)^F\} =\{\overline{\chi}^\alpha_m,(-1)^F\} = 0,\\
    &(-1)^F\ket{0} =\ket{0},\quad (-1)^F\ket{-\tfrac{1}{8}} =\ket{-\tfrac{1}{8}},\quad  (-1)^F\ket{\Omega}=\ket{\Omega}.
\end{split}
\end{equation}
We define traces and super-traces, that is traces with and without a $(-1)^F$ insertion, in the case of the irreducible symplectic fermion modules,
\begin{equation}
    \mathrm{Tr}_{\lambda,1/2}(\mathcal{O}) = \mathrm{Tr}_{\mathbb{L}_\lambda}(\mathcal{O}), 
    \qquad 
    \mathrm{Tr}_{\lambda,0}(\mathcal{O}) = \mathrm{Tr}_{\mathbb{L}_\lambda}\big((-1)^F\mathcal{O}\big).
\end{equation}
The corresponding characters and super-characters are
\begin{equation}
    \chi_{\lambda,s}(\t) 
    = \mathrm{Tr}_{\lambda,s}(q^{L_0-c/24}) 
    = q^{h(\lambda)+\tfrac{1}{12}} 
      \prod_{n \in \mathbb{N}-\lambda} \big(1 - e^{2\pi i s}q^n\big)^2,
\end{equation}
with $h(0)=0$ and $h(1/2)=-1/8$ (and in generic twisted sectors $h(\lambda) = \lambda(\lambda-1)/2$). We have also defined $q = \exp(2\pi i \t)$. 

Using these expressions, we can write the traces over the spaces \eqref{eq: Even Odd Decompositions} as
\begin{align}
\begin{split}
    \Tr_{\SCW_0}(q^{L_0+1/12)}) &= 
    \tfrac 12 ( \chi_{0,1/2} + \chi_{0,0})
    \;,\;\;
\Tr_{\SCW_1}(q^{L_0+1/12)}) = 
    \tfrac 12 ( \chi_{0,1/2} - \chi_{0,0})
    \;,\;\;\\
    \Tr_{\SCW_{-1/8}}(q^{L_0+1/12)}) &= 
    \tfrac 12 ( \chi_{1/2,1/2} + \chi_{1/2,0})
    \;,\;\;
\Tr_{\SCW_{3/8}}(q^{L_0+1/12)}) = 
    \tfrac 12 ( \chi_{1/2,1/2} - \chi_{1/2,0})
    \;.\;\;
\end{split}
\end{align}

We find that the contribution to the modular invariant from $(\SCW_{-1/8}\otimes\overline{\SCW_{-1/8}})\oplus (\SCW_{3/8}\otimes\overline{\SCW_{3/8}})$, the irreducible part of the state space \eqref{eq: Triplet state space}, is
\begin{equation}
    \tfrac{1}{2}\,|\chi_{1/2,1/2}|^2
    + \tfrac{1}{2}\,|\chi_{1/2,0}|^2
\end{equation}
When we consider $\SCB$, $L_0$ and $\bar L_0$ are diagonalisable on $V^o_{\ket{\psi^+}}\oplus V^o_{\ket{\psi^-}}$, but on $V^e_{\ket\Omega}\oplus V^e_{\ket 0}$, they are not. On the basis states 
\begin{align}
 \begin{pmatrix}
 \chi^-_{-n_1}...\chi^-_{-n_{\nu}}\chi^+_{-m_1}...\chi^+_{-m_{\mu}}\overline{\chi}^-_{-\overline{n}_1}...\overline{\chi}^-_{-{\overline{n}}_{\sigma}}\overline{\chi}^+_{-\overline{m}_1}...\overline{\chi}^+_{-\overline{m}_{\rho}}\ket{0}
 \\
 \chi^-_{-n_1}...\chi^-_{-n_{\nu}}\chi^+_{-m_1}...\chi^+_{-m_{\mu}}\overline{\chi}^-_{-\overline{n}_1}...\overline{\chi}^-_{-{\overline{n}}_{\sigma}}\overline{\chi}^+_{-\overline{m}_1}...\overline{\chi}^+_{-\overline{m}_{\rho}}\ket{\Omega}
 \end{pmatrix}    
\;,
 \end{align}
with 
\begin{align}
N = \sum_i n_i \,{+} \sum_j m_j
    \;,\;\;
    \bar N = \sum_i \bar n_i \,{+} \sum_j \bar m_j
    \;,
\end{align}
they take the form
\begin{align}
    L_0 =
    \begin{pmatrix}
    N & 0 \\ 1 & N    
    \end{pmatrix}
    \;,\;\;\;\;
    \bar L_0 =
    \begin{pmatrix}
    \bar N & 0 \\ 1 & \bar N    
    \end{pmatrix}
    \;.\;\;
\end{align}
Using $q^a = e^{a \log q}$, we get 
\begin{align}
    q^{L_0} = \begin{pmatrix}
        q^N & 0 \\ \log(q)\, q^N\;\; & q^N
    \end{pmatrix}
    \;,\;\;\;\;
    q^{\bar L_0} = \begin{pmatrix}
        q^{\bar N} & 0 \\ \log(q) \,q^{\bar N }\;\;& q^{\bar N}
    \end{pmatrix}
    \;.
\end{align}
and the off-diagonal terms do not contribute to any traces.
Hence, using a combination of traces and super-traces in the bulk module $\bb{S}_0\otimes \overline{\bb{S}_0}$ with the relations \eqref{eq: Bulk Ground state relations}, one can see that the contribution of the bulk triplet module $\SCB$ to the modular invariant is
\begin{equation}
    \Tr_\SCB(q^{L_0-c/24}\overline{q}^{\overline{L}_0-c/24}) = 2\, |\Tr_{\bb{L}_0}(q^{L_0-c/24})|^2 = 2 |\chi_{0,1/2}|^2.
\end{equation}
We emphasise now that the form of this relation will be unchanged when considering GGEs later. Looking at the expression for charges in the bilinear hierarchy 
\eqref{eq: THE CHARGES}, one sees that they commute with $L_0$ and $\bar L_0$ and furthermore are diagonalisable, and so 
\begin{equation}
    \Tr_\SCB(q^{L_0-c/24}\overline{q}^{\overline{L}_0-c/24}
    e^{\a \Q_n + \a' \overline{\Q}_n}) 
    = 2 
    \Tr_{\bb{L}_0}(q^{L_0-c/24}e^{\a\Q_n})
    \times
    \Tr_{\bb{L}_0}(\bar q^{L_0-c/24}e^{{\a}'\overline \Q_n})
\end{equation}

Piecing this all together tells us that the modular invariant partition function of the triplet model is
\begin{equation}\label{eq: triplet Modular Invariant.}
    Z_{\text{triplet}}(\tau,\bar{\tau})
    = \tfrac{1}{2}\,|\chi_{1/2,1/2}|^2
    + \tfrac{1}{2}\,|\chi_{1/2,0}|^2
    + 2\,|\chi_{0,1/2}|^2.
\end{equation}
This is a well-defined partition function, and will serve as the basis for a GGE. It is clear from the discussion that we need only consider the case of the traces of the modules $\bb{L}_0$ and $\bb{L}_{1/2}$.

One can indeed check that \eqref{eq: triplet Modular Invariant.} is a modular invariant. This follows from the modular properties of the characters and super-characters of the irreducible symplectic fermion modules $\bb L_0$ and $\bb L_{1/2}$ which we elaborate here.

The action of the generators of the modular group on the modular parameter are $T: \tau \mapsto \tau+1$, $S:\tau \mapsto -1/\t$. The characters $\chi_{0,1/2}, \chi_{1/2,1/2}$ and $\chi_{1/2,0}$ transform in the three dimensional representation
\begin{align}
        T:&  \begin{pmatrix}
        \chi_{0,1/2}(\tau+1)
        \\
        \chi_{1/2,1/2}(\tau+1)
        \\
        \chi_{1/2,0}(\tau+1)
        \end{pmatrix} 
        = 
        \begin{pmatrix}
        e^{\frac{i \pi}{6}} & 0 & 0 \\
        0 & e^{-\frac{i \pi}{12}} & 0 \\
        0 & 0 & e^{-\frac{i \pi}{12}} \\
        \end{pmatrix}
        \begin{pmatrix}
        \chi_{0,1/2}(\tau)\\
        \chi_{1/2,1/2}(\tau)\\
        \chi_{1/2,0}(\tau)
    \end{pmatrix}
    \label{eq: T trasform}
\;,\;\;\;\;\\
S:&  \begin{pmatrix}
        \chi_{0,1/2}(-1/\tau)
        \\
        \chi_{1/2,1/2}(-1/\tau)
        \\
        \chi_{1/2,0}(-1/\tau)
        \end{pmatrix} 
        = 
          \begin{pmatrix}
        ~0 & ~~0~~ & \tfrac{1}{2}\\
        ~0 & 1 & ~0~\\
        ~2 & 0 & ~0~
    \end{pmatrix}
    \begin{pmatrix}
        \chi_{0,1/2}(\tau)\\
        \chi_{1/2,1/2}(\tau)\\
        \chi_{1/2,0}(\tau)
    \end{pmatrix}
    \label{eq: S trasform}
    \;.
\end{align}
The character $\chi_{0,0}$ however transforms as 
\begin{align}
    \chi_{0,0}(\tau+1) = e^{i\pi/6}\chi_{0,0}(\tau)
    \;,\;\;
    \chi_{0,0}(-1/\tau) = - i \tau \chi_{0,0}(\tau)\;.
\end{align}
These are closed as functional equations but to get a proper linear representation of the modular group one must consider in addition the pseudo-trace $g(\tau) = \tau \chi_{0,0}(\tau)$ \cite{Gainutdinov:2016qhz}:
\begin{align}
        T:&  \begin{pmatrix}
        \chi_{0,0}(\tau+1)
        \\
        g(\tau+1)
        \end{pmatrix} 
        = 
        e^{\frac{i \pi}{6}}\begin{pmatrix}
        ~1~~ & 0~ \\
        ~1~~ & 1~
        \end{pmatrix}
        \begin{pmatrix}
        \chi_{0,0}(\tau)\\
        g(\tau)
    \end{pmatrix}
    \label{eq: T pseudo character}
\;,\;\;\;\;\\
S:&  \begin{pmatrix}
        \chi_{0,0}(-1/\tau)
        \\
        g(-1/\tau)
        \end{pmatrix} 
        = 
          \,i\,\begin{pmatrix}
          ~~0 & ~~1 ~ \\
          -1  & ~~0~
    \end{pmatrix}
    \begin{pmatrix}
        \chi_{0,0}(\tau)\\
        g(\tau)
    \end{pmatrix}
    \label{eq: S pseudo character}
    \;.
\end{align}
Our goal will be to extend these characters with additional conserved charges, and understand the modular properties of those extended characters, in particular the modular $S$-transformation.

\section{Hierarchy of Bilinear Fields and Charges}\label{sec: Hierarchy of Bilinear Fields}

It has long been known that one can construct an infinite tower of bosonic fields which are bilinear in the symplectic fermions \cite{Shen:1992dd}.
 Since these fields are bilinear, their zero modes act linearly on the modes of the symplectic fermions and hence the zero modes all mutually commute. 
 
In this paper, we are interested not so much in the zero modes of the bilinear fields, as in the corresponding charges which are the expressions for the integrals of the bilinear fields around a cylinder, which requires us to map them to the complex plane using the method of \cite{Gaberdiel:1994fs}.
In this section we recall the form of the explicit basis of quasi-primary bilinear fields
and find the explicit expressions for the corresponding charges. 

The charges are easily determined up to the addition of a constant, and we show how one can fix this constant in two separate ways.

\subsection{Quasi-Primary Bilinear Fields}
\label{subsec:basisoffields}

An infinite set of quasi-primary fields of all integer weights greater than or equal to two was already found at $c=-2$ in \cite{Shen:1992dd},
where they were identified as the generators of
the ``linear'' $W_{\infty}$ algebra.
The fields are realised in \cite{Shen:1992dd} as a set of quasi-primary fields which are bilinear in the fields of a $bc$ ghost system. The weight 1 and weight 0 ghost fields were denoted $\psi$ and $\bar\psi$ respectively, but the fields of the $W_\infty$ algebra are expressed only in terms of $\psi$ and $\partial\bar\psi$, which can then be identified with the symplectic fermion fields $\chi^+$ and $\chi^-$.
Rewriting the results in \cite{Shen:1992dd} for later convenience, we find that the quasi-primary bilinear field of weight $n$ is 
\begin{equation}\label{eq: Bilinear field, but Ank}
    B_n(z)=\sum_{k=0}^{n-2}A_{n,k}(\del^{n-2-k}\chi^- \del^k \chi^+)(z),
\end{equation}
where
\begin{equation}\label{eq: A n k}
    A_{n,k}=\left[\frac{1}{(n-2)!}\binom{2n-2}{n-1} \right]^{-1}\binom{n}{k+1}  \frac{(-1)^k }{(n-2-k)!k!}.
\end{equation}
One can simplify this expression by collecting most of the terms into a total derivative
\begin{equation}\label{eq: Bilinear Fields}
    B_n(z) = (\del^{n-2}\chi^- \chi^+)(z) \text{ mod }\del, \quad n =2,3,4,...
\end{equation}
Here,$\text{ mod }\del$ means that we have dropped total derivative terms. One can easily check that $B_n(z)$ is a quasi-primary field by calculating
\begin{equation}
L_1 \ket{B_{n}}=0.
\end{equation}
where $\ket{B_n}$ is the state associated to the field $B_n(z)$. We now turn to the charges associated to the fields $B_n$. 

\subsection{Charges associated to quasi-primary fields}

\begin{figure}[htb]
\[
\begin{array}{cccccccccc}
&&
\hfill\mbox{\Large$\llcorner$}\raisebox{1mm}{\kern -1.4mm{$w$}}
&&
\hfill\mbox{\Large$\llcorner$}\raisebox{1mm}{\kern -1.4mm{$z$}}
\\[-3mm]
\begin{tikzpicture}[baseline=2em]
\begin{scope}[yshift=1.4em]
\draw (0,0.2) -- (2.,0.2);
\draw (0,1.2) -- (2.,1.2);
\draw (0,0.7) ellipse (0.2 and 0.5);
\draw (2.,0.2) arc (-90:90:0.2 and 0.5);
\draw[dashed] (2.,1.2) arc (90:270:0.2 and 0.5);
\draw[->,green] (0.2,.5) -- (1.4,0.5);
\draw[red] (0.8,0.7) -- (0.9,0.8) -- (1.0,0.7);
\draw[red] (0.7,0.2) arc (-90:90:0.2 and 0.5);
\draw[dashed,red] (.7,1.2) arc (90:270:0.2 and 0.5);
\draw[->] (-0.48,0.35) arc (-120:150:0.16 and 0.4) node [above left] {$\scriptstyle R$};
\end{scope}
\end{tikzpicture}
&
\raisebox{4mm}{\hbox{$=$}}
&
\begin{tikzpicture}[baseline=2em]
\begin{scope}[yshift=1.2em]
\draw[->,green] (0.2,.5) -- (1.4,0.5);
\draw (0,0.2) -- (2.,0.2);
\draw (2,1.4) -- (0,1.4);
\draw (1.4,.1) -- (1.2,.3);
\draw (1.4,1.3) -- (1.2,1.5);
\draw[red] (0.7,0.2) -- (0.7,1.4);
\draw[red] (0.6,0.7) -- (0.7,0.8) -- (0.8,0.7);
\draw[<->] (-0.2,0.2) -- (-0.2,1.4);
\node at (-0.4,0.8) {$\scriptstyle 2\pi$} ;
\end{scope}
\end{tikzpicture}
&
\raisebox{4mm}{\hbox{$\equiv$}}
&
\begin{tikzpicture}[baseline=2em]
\begin{scope}[yshift=3em]
\draw[->,green] (0.,0.) -- (0.7,1.17);
\draw (-1.4,0)--(1.4,0);
\draw (0,-1.4)--(0,1.4);
\draw[red] (0.565,0.405) -- (0.565,0.565) -- (0.725,0.565);
\draw[red] (0,0) circle (0.8);
\end{scope}
\end{tikzpicture}
\\[4mm]
  && \displaystyle\int_0^{2\pi i} \frac{dw}{2\pi i}J_n(w)
  && \displaystyle \oint \frac{dz}{2\pi i z} [z^{L_0} \!\cdot \!\hat J_n](z)
\\[3mm]
(a) && (b) && (c) 
\end{array}
\]
\caption{Illustration of (a) the current $J_n$ being integrated on a fixed spatial slice of an infinitely long cylinder being equivalent to (b) a line integral in the plane of $J_n(w)$ and (c) a contour integral of a local field $[z^{L_0} \cdot \hat{J}_n](z)$ on the plane. The green arrow indicates the time direction.}
    \label{fig: Cyl to Plane}
\end{figure}
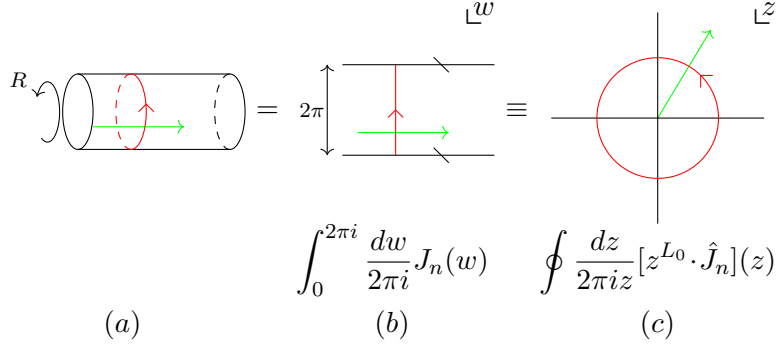

Suppose that $J(w)$ is a quasi-primary field with conformal weight $n$. The charge $I$ associated to $J(w)$ is obtained by integrating the field on a fixed spatial slice of a cylinder. 
It will be convenient to take $w=2\pi u/R$ so that the cylinder has 
circumference $2\pi$ and is given by the strip in the complex plane $0\leq\text{Im}(w)<2\pi$ with periodic boundary conditions in the imaginary direction.
We then have 
\begin{equation}
I = 
    \int_0^{2\pi i}\frac{dw}{2\pi i}J(w).
\end{equation}
We can express $J(w)$ in terms of fields on the plane by mapping from the cylinder with coordinate $w$ to the complex plane with the map $z=e^{2\pi u/R} = e^w$. 
The result, using, for example, the technique in \cite{Gaberdiel:1994fs}, is 
\begin{equation}\label{eq: the hats}
    J(w) = [z^{L_0} \cdot \hat{J}](z),
\end{equation}
where $\hat{J}(z)$ is a sum of local fields on the plane with weight up to and including $n$.
It will be important for us later that the leading term in $\hat J$ is the field $J$ itself, all other contributions being of lower weight.
The hat notation is introduced in appendix I of \cite{Downing:2025huv} to avoid clutter with many powers of $z$ appearing. 
The notation for the action $L_0\cdot$  on a local field in \eqref{eq: the hats} is a shorthand for
\begin{equation}
    [L_0 \cdot \varphi](z) = V(L_0 \ket{\varphi},z)\;,
\end{equation}
where we have used the vertex operator notation, $V(\ket{\phi},z)$, denoting the local field that corresponds to the state $\ket{\phi}$.
Note that
the charge on the cylinder 
is given by
\begin{equation} \label{eq: I = hat J 0}
I =    \int_0^{2\pi i}\frac{dw}{2\pi i}J(w) = \oint \frac{dz}{2\pi i z} \,[z^{L_0}\cdot\hat{J}](z)
=  \hat J_0 
\;.
\end{equation}
and will amount to being a sum of zero-modes of fields on the plane, including a possible constant term. 
See Figure \ref{fig: Cyl to Plane} for a illustration of this process.

An illustrative example of the above is  given by the quasi-primary field $J(w) = \Lambda(w) = (TT)(w)-\frac{3}{10}T''(w)$ where $T$ is the stress tensor. Here we have:
\begin{equation}
\begin{split}
    \hat{\Lambda}(z) &= \Lambda(z) - \frac{5c+22}{60}T(z) + \frac{c(5c+22)}{2880},\\
    \Lambda(w)&= [z^{L_0}\cdot \hat{\Lambda}](z) = z^4\Lambda(z) - \frac{5c+22}{60}z^2T(z) + \frac{c(5c+22)}{2880},\\
    \hat{\Lambda}_0 & = \oint \frac{dz}{2\pi i z}\left(z^4\Lambda(z) - \frac{5c+22}{60}z^2T(z) + \frac{c(5c+22)}{2880}\right)\\
    &= \Lambda_0 - \frac{5c+22}{60}L_0 + \frac{c(5c+22)}{2880}.
\end{split}
\end{equation}
The planar zero-mode of $\Lambda(z)$ is $\Lambda_0$, but the cylinder zero-mode of $\Lambda(w)$ is $\Lambda_0 - \frac{5c+22}{60}L_0 + \frac{c(5c+22)}{2880}$, a sum of three planar zero-modes.

\subsection{Expressions for the Charges in the Bilinear Hierarchy}\label{subsec: Expressions for the charges}

We now apply the previous section to the currents $B_n$. Our first aim is to find explicit expressions for the associated charges,
\begin{equation}\label{eq: Q = int B}
    \Q_{n-1} = (\hat{B}_n)_0 = \int_0^{2\pi i}\frac{dw}{2\pi i}B_n(w)
    \;.
\end{equation}
We will do this in two ways: firstly in section 
\ref{sec:theway} using means of modular forms, and then in section 
\ref{sec:zeta} by means of zeta-function regularisation.

\subsubsection{The Way of Modular Forms}
\label{sec:theway}
Reviews of modular forms can be found in many places; we will use the conventions of the appendix of \cite{Downing:2021mfw}, which we repeat in appendix \ref{app: Modular Forms} for convenience.

In  \cite{Downing:2025huv,Dijkgraaf:1996iy}, it is argued that if  
$\{J_i\}$ are a set of quasi-primary fields of weights $n_i$, and the zero modes 
$\{(\hat{J}_{i})_0\}$ 
are mutually commuting, then 
\begin{equation}\label{eq: vev jjjjj}
\vev{(\hat{J}_{i_1})_0(\hat{J}_{i_2})_0...(\hat{J}_{i_N})_0},
\end{equation}
is a quasi-modular form of weight $\sum_in_i$ and depth $N-1$.

We can use this fact to fix the form of the charges $\Q_{n-1}$ as follows. Since we know the explicit form of $B_n$, we know both the explicit forms of the leading term of $\hat B_n$ and of $\Q_{n-1}=(\hat B_n)_0$ and the general form of the sub-leading terms. Requiring the correct modular properties will fix the sub-leading terms uniquely.

The map from $B_n(w)$ to $\hat B_n(z)$ only uses the action of the Virasoro algebra positive modes on the state $|B_n\rangle$ and so the result will again be a bilinear field in the symplectic fermion modes (possibly with an additional constant term), but this time a sum over fields of all weights up to $n$, of the form
\begin{equation}\label{eq:Bhatnansatz}
    \hat{B}_{n}(z) = (\del^{n-2}\chi^- \chi^+)(z)
    +\sum_{k=2}^{n-1} C^n_{k}(\del^{k-2}\chi^- \chi^+)(z) +  C^n_0 + \del(\ldots).
\end{equation}
We can further restrict this form by considering the symmetry of the symplectic fermion algebra,
\begin{align}
    \chi^- \to \chi^+\;,\;\; \chi^+ \to - \chi^-
    \;.
\end{align}
The term $\del^n \chi^- \chi^+ + \del(\ldots)$ picks up a sign $(-1)^n$ under this symmetry, and so 
$B_n(z) \mapsto (-1)^n B_n(z)$ while
the Virasoro algebra is invariant, 
so the only terms possible in \eqref{eq:Bhatnansatz} are those with the same parity under the symmetry, and we can refine the answer to
\begin{align}
    \hat{B}_{2n}(z) &= (\del^{2n-2}\chi^- \chi^+)(z)
    +\sum_{k=1}^{n-1} C^{2n}_{2k}(\del^{2k-2}\chi^- \chi^+)(z) +  C^{2n}_0 + \del(\ldots)
    \;,
    \label{eq:Bhatnansatz2}
    \\
        \hat{B}_{2n+1}(z) &= (\del^{2n-1}\chi^- \chi^+)(z)
    +\sum_{k=1}^{n-1} C^{2n+1}_{2k+1}(\del^{2k-1}\chi^- \chi^+)(z) +  \del(\ldots).
    \label{eq:Bhatnansatz2odd}
\end{align}
Note that there is no constant term in \eqref{eq:Bhatnansatz2odd}.
When we take the zero mode of $\hat B_n$, we then find
\begin{align}
    \Q_{2n-1} = (\hat{B}_{2n})_0 &= 
    \sum_{j \in \bb{Z}+\lambda}\Big( j^{2n-2}  
    +\sum_{k=1}^{n-1} H^{2n}_{2k}\, j^{2k-2} \Big)
    : \chi^-_{-j}\chi^+_j:
    \;+\;H^{2n}_0
    \;,
\label{eq:Bhatnansatz3}
    \\
        \Q_{2n}= (\hat{B}_{2n+1})_0 &= 
    \sum_{j \in \bb{Z}+\lambda} \Big( j^{2n-1}  
    +\sum_{k=1}^{n-1} H^{2n+1}_{2k+1}\, j^{2k-1} \Big)
    : \chi^-_{-j}\chi^+_j:\;,
\label{eq:Bhatnansatz3odd}
\\
        \Q_{2}= (\hat{B}_{3})_0 &= 
    \sum_{j \in \bb{Z}+\lambda}  j^{}  
    : \chi^-_{-j}\chi^+_j:\;,
\label{eq:Bhatnansatz3Q2}
\end{align}
where $:\cdots:$ is normal ordering and we adopt the convention that
\begin{equation}
    :\!\chi^-_0\chi^+_{0}\!\!:\;\; =
    - :\!\chi^+_0\chi^-_0\!\!:\;\; = \chi^-_0\chi^+_{0}.
\end{equation}
We could, in principle, calculate the constants $C^n_k$ and $H^n_k$ directly, but it is much easier to use the modular properties of the $n$-point functions of the charges, as these determine them uniquely. 
This argument follows very much the lines in \cite{Downing:2021mfw}, but with one extra important subtlety.
For odd weight charges, 
we know that the torus one-point functions $\vev{\Q_{n}}$ are modular forms and these can be calculated as traces of the symplectic fermion Hilbert space and this fixes the charges uniquely once the leading term is known.
However, the traces 
$ \vev{\Q_{n}}$ vanish identically for even spin charges and we have to consider not only these one-point functions, but the traces $\vev{\Q_{n_1}\Q_{n_2}}$ as well.

We consider the cases of fields of even and odd weight, with charges of odd and even weight respectively in the next two sections.

\subsubsection{Even Weight Fields and Odd Spin Charges}

As is argued in, for example, \cite{Downing:2021mfw}, the torus one point function of zero-modes of quasi-primary fields 
are 
modular forms. These one-point functions are written in terms of traces over the Hilbert space of the theory which is expressed as traces over the representations 
$\bb{L}_\lambda$. 

To this end, 
we will now introduce a shorthand notation for these traces.
We define
\begin{equation}\label{eq: Def Exp Value +}
    \vev{\,\CO\,}_{\lambda,s}(\t) 
    = \Tr_{\lambda,s}(\,\CO\, q^{L_0 - c/24})\,\;,
\end{equation}
with 
\begin{equation}
    q = e^{2\pi i \tau}\;,\quad\quad \lambda,s \in \{0 , \frac{1}{2}\}\;.
\end{equation}
In this notation, the characters are 
$\vev{1}_{\lambda,s}$.

Following the arguments of section \ref{ssec:triplet}, the one point function of a chiral operator $\CO$ is then given as 
\begin{align}
    \langle\,\CO\,\rangle
    = 
    \tfrac 12
    \vev{\,\CO\,}_{1/2,1/2} \bar\chi_{1/2,1/2}
    + 
    \tfrac 12
    \vev{\,\CO\,}_{1/2,0} \bar\chi_{1/2,0}
    + 
    2
    \vev{\,\CO\,}_{0,1/2} \bar\chi_{0,1/2}
    \;.
\end{align}
For this to transform as a weight $n$ modular form, we need
the individual traces to transform in the same way as the characters, up to a factor of $\tau^{2n}$, that is (including the sector $(\lambda,s)=(0,0)$ for completeness),
\begin{equation}\left\langle
\,\CO\,
\right\rangle_{\lambda,s}(-\tfrac{1}{\t}) = \t^{2n}
\,\frac{2^{\delta_{s,0}\delta_{\lambda,1/2}}}{2^{\delta_{s,1/2}\delta_{\lambda,0}}}
      \left(\frac{\tau}{i}\right)^{\delta_{\lambda,0}\delta_{s,0}}
      \left\langle 
\,\CO\,
\right\rangle_{s,\lambda}(\t).
\end{equation}
Note that we have packaged the S-transformations of all the sectors \eqref{eq: T trasform} \eqref{eq: S trasform} \eqref{eq: T pseudo character} \eqref{eq: S pseudo character} into one expression.
This has been done so that when we look at the exact transformation of the GGE later, we can treat all sectors at once. It is important to remember that only three of these form a linear representation, and that $\lambda = s =0$ belongs to a slightly different representation \eqref{eq: T pseudo character}, \eqref{eq: S pseudo character}.

Now denote by $\Q'_{2k-1}$ just the single piece appearing in the ansatz for the charge \eqref{eq:Bhatnansatz3}:
\begin{equation}
    \Q'_{2k-1} = \sum_{m\in \bb{Z}+\lambda} j^{2k-2}:\chi^-_j\chi^+_{-j}:
    \;.
\end{equation}
We have 
\begin{equation}
    [\Q'_{2k-1},\chi^{\pm}_n] = - n^{2k-1}\chi^{\pm}_n,
\end{equation}
so that:
\begin{equation}
    \vev{e^{\mu\Q'_{2k-1}}}_{\lambda,s} = q^{h(\lambda)+\frac{1}{12}}\prod_{n=1}^{\infty}(1-e^{2\pi i s} q^{n-\lambda}e^{\mu{(n-\lambda)}^{2k-1}})^2.
\end{equation}
Then we can determine $\vev{\Q'_{2k-1}}$ by:
\begin{equation}
\vev{\Q'_{2k-1}}_{\lambda,s} = \frac{d}{d\mu}\biggl |_{\mu =0}   \vev{e^{\mu\Q'_{2k-1}}}_{\lambda,s}.
\end{equation}
Doing this calculation yields:
\begin{equation}
\vev{\Q'_{2k-1}}_{\lambda,s} =  -e^{2\pi i s} 2\left[\sum_{m=1}^{\infty}\left( \frac{(m-\lambda)^{2k-1}q^{m-\lambda}}{1-e^{2\pi is} q^{m-\lambda}}\right)\right]  \vev{1}_{\lambda,s}
\end{equation}
Looking at each case in turn, we find that we can write the prefactor in front of $\vev{1}$ in terms of the Eisenstein series.
\begin{align}
\vev{\Q'_{2k-1}}_{0,\frac{1}{2}}
& =\zeta(1-2k)\big[\left(E_{2k}(\t) - 2 E_{2k}(2\t )\right) + 1\big]\vev{1}_{0,1/2}
\;,
\\
\vev{\Q'_{2k-1}}_{\frac{1}{2},\frac{1}{2}} 
&= \zeta(1-2k)\Big[\frac{1}{2^{2k-1}}\left(E_{2k}(\tfrac{\t}{2}) + 2^{2k}E_{2k}(2 \t) - (2^{2k-1}+2)E_{2k}(\t)\right) 
\nonumber\\
&\qquad\qquad +\left(\left(\tfrac{1}{2}\right)^{2k-1}-1\right)\Big] \vev{1}_{\frac{1}{2},\frac{1}{2}}
\;,
\\
\vev{\Q'_{2k-1}}_{\frac{1}{2},0}
&= \zeta(1-2k)\left[\frac{1}{2^{2k-1}}\left(2^{2k-1}E_{2k}(\t)-E_{2k}(\tfrac{\t}{2})\right) + \left(\left(\tfrac{1}{2}\right)^{2k-1}-1\right)\right] \vev{1}_{\frac{1}{2},0}
\;,
\\
\vev{\Q_{2k-1}'}_{0,0} &= \zeta(1-2k)(1-E_{2k}(\t))\vev{1}_{0,0}
\;.
\end{align}
In each case, we have a weight $2n$ modular form (under the group $\Gamma(2)$ (defined in appendix \ref{app: Modular Forms}) which we shall denote by $\CF_{2k}^{\lambda,s}(\t)$ and some additional constant given by:
\begin{equation}
    {\bf c}_{2k-1}(\lambda) = \begin{cases}
         \zeta(1-2k)& \lambda =0,\\
         \left(\left(\tfrac{1}{2}\right)^{2k-1}-1\right)\zeta(1-2k)& \lambda = 1/2.
    \end{cases}
\end{equation}
In essence, we find:
\begin{equation}
    \vev{\Q'_{2k-1}}_{\lambda,s} = \left(\CF_{2k}^{\lambda,s} + {\bf c}_{2k-1}(\lambda)\right)\vev{1}_{\lambda,s}
\end{equation}
Plugging these results into \eqref{eq:Bhatnansatz3}, we find:
\begin{equation}\label{eq: Odd-Spin One point function}
\begin{split}
    \vev{\Q_{2n-1}}_{\lambda,s} 
    &=\vev{\Q'_{2n-1}}_{\lambda,s} 
    + \sum_{k=1}^{n-1}H^{2n}_{2k}\vev{\Q'_{2k-1}}_{\lambda,s} 
    + \vev{H_0}_{\lambda,s} \\
    &=\left[H_0 + \CF_{2n}^{\lambda,s} + {\bf c}_{2n-1}(\lambda) 
    + \sum_{k=1}^{n-1}H^{2n}_{2k}\left(\CF_{2k}^{\lambda,s} 
    + {\bf c}_{2k-1}(\lambda)\right)\right] \vev{1}_{\lambda,s}.\\
\end{split}
\end{equation}
This is a weight $2n$ vector valued modular form if and only if:
\begin{equation}
    H_{k>0} = 0 \quad,\quad H_0 = - {\bf c}_{2n-1}(\lambda).
\end{equation}
So the final result is that the charge $\Q_{2n-1}$ is:
\begin{equation}\label{Odd Spin Charge}
    \Q_{2n-1} = \sum_{j \in \bb{Z}+\lambda}j^{2n-2}:\chi^-_j\chi^+_{-j}:  - {\bf c}_{2n-1}(\lambda).
\end{equation}
and
\begin{equation}\label{Odd Spin Charge FORM}
    \vev{\Q_{2n-1}}_{\lambda,s} = \CF_{2n}^{\lambda,s}\vev{1}_{\lambda,s}.
\end{equation}

\subsubsection{Odd Weight Fields and Even Spin Charges}
We would like to repeat the above process for the odd weight fields, which give even spin charges:
\begin{equation}\label{eq: Q 2n ansatz}
    \Q_{2n} = (\hat{B}_{2n+1})_0
\end{equation}
We define now the auxiliary expression similarly to the previous section,
\begin{equation}
    \Q'_{2k} =  \sum_{j\in \bb{Z}+\lambda} j^{2k-1}:\chi^-_{-j}\chi^+_{j}:
    \;,
\end{equation}
(from \eqref{eq:Bhatnansatz3Q2} it is obvious that $\Q_2=\Q_2'$). One can verify that:
\begin{equation}
    [\Q'_{2k},\chi^\pm_n] = \mp n^{2k}\chi^{\pm}_n \;.
\end{equation}
Using this, one can calculate the expectation value:
\begin{equation}
    \vev{e^{\mu\Q'_{2k}}}_{\lambda,s} = q^{\frac{\lambda(\lambda-1)}{2}+\frac{1}{12}}\prod_{n=1}^{\infty}(1-e^{2\pi i s} q^{n-\lambda}e^{\mu{(n-\lambda)}^{2k}})(1-e^{2\pi is} q^{n-\lambda}e^{-{\mu(n-\lambda)}^{2k}}).
\end{equation}
Then we find that the one-point function of the even spin auxiliary charge vanishes:
\begin{equation}
    \vev{\Q'_{2k}}_{\lambda,s} = \frac{d}{d\mu}\biggl|_{\mu=0}  \vev{e^{\mu\Q'_{2k}}}_{\lambda,s} = 0.
\end{equation}
Since the one point functions do not give a useful result,
we now demand, instead, that $ \vev{\Q_{2n}\Q_2}$ is a weight $2n+4$ depth $1$ quasi-modular form.
This fixes the coefficients $H^{2n+1}_{2k+1}$ as follows. 

We need to consider the pieces $\vev{\Q'_{2k}\Q_2}$.
We have the following trace:
\begin{equation}
\begin{split}
    \vev{e^{\mu\Q'_{2k} + \nu\Q_2}}_{\lambda,s} = q^{h(\lambda)+\frac{1}{12}}\prod_{n=1}^{\infty}&(1-e^{2\pi i s} q^{n-\lambda}e^{\mu{(n-\lambda)}^{2k}+ \nu(n-\lambda)^2})\\
    &\quad\times(1-e^{2\pi i s} q^{n-\lambda}e^{-{\mu(n-\lambda)}^{2k} - \nu(n-\lambda)^2}).
\end{split}
\end{equation}
We can obtain our desired two-point function by taking derivatives as before:
\begin{equation}
\begin{split}
\vev{\Q'_{2k}\Q_2}_{\lambda,s} &= \frac{\del^2}{\del \mu \,\del\nu} \biggl|_{\mu,\nu =0}\vev{e^{\mu\Q'_{2k} + \nu\Q_2}}_{\lambda,\pm}\\
&= -e^{2\pi i s} 2 \sum_{n=1}^{\infty}\left[\frac{(n-\lambda)^{2k+2}q^{n-\lambda}}{1-e^{2\pi i s} q^{n-\lambda}} +e^{2\pi is }\frac{(n-\lambda)^{2k+2}q^{2n-2\lambda}}{(1-e^{2\pi is} q^{n-\lambda})^2}\right] \vev{1}_{\lambda,s}\\
\end{split}.
\end{equation}
Recall from Appendix \ref{app: Modular Forms},
\begin{equation}
    D_r = q\frac{\del}{\del q} - \frac{r}{12}E_2(q) = \frac{1}{2\pi i}\frac{d}{d\t} - \frac{r}{12}E_2(\t),
\end{equation}
the covariant Serre derivative. This has the properties that if $F_r$ denotes a weight $r$ modular form, then  $D_r F_r$ is a weight $r+2$ modular form. We can see by definition that $D_0F_r$ is also a weight $r+2$ and depth $1$ quasi-modular form. We can break the two-point function down to each sector and rewrite them in terms of Eisenstein series.
\begin{align}\label{eq: two point of evens}
\vev{\Q'_{2k}\Q_{2}}_{0,\frac{1}{2}}&=\zeta(1-2(k+1))\left[D_0 E_{2(k+1)}(\t) - 4 D_0 E_{2(k+1)}(2\t)\right]\vev{1}_{0,\frac{1}{2}}, \\
\vev{\Q'_{2k}\Q_{2}}_{\frac{1}{2},\frac{1}{2}}&= \zeta(1-2(k+1))\biggl[(4^{-k-1}D_0 E_{2(k+1)}(\tfrac{\t}{2}) \nonumber\\
&\quad -2^{-2 k} \left(2^{2 k}+1\right)D_0E_{2(k+1)}(\t) + 4 D_0 E_{2(k+1)}(2\t)\biggr]\vev{1}_{\frac{1}{2},\frac{1}{2}},\\
\vev{\Q'_{2k}\Q_{2}}_{\frac{1}{2},0} & = \zeta(1-2(k+1))\left[-4^{-k-1}D_0 E_{2(k+1)}(\tfrac{\t}{2})+D_0E_{2(k+1)}(\t)\right]\vev{1}_{\frac{1}{2},0},\\
\vev{\Q'_{2k}\Q_{2}}_{0,0} &= - \zeta(1-2(k+1))D_0 E_{2(k+1)}(\t)\vev{1}_{0,0} .
\end{align}
Now we have shown that $\vev{\Q'_{2k}\Q_2}_{\lambda,s}$ forms a weight $2k+4$ depth $1$ vector-valued quasi-modular form. Given that we know
\begin{equation}
    \vev{\Q_{2n}\Q_2} = \vev{\Q'_{2n}\Q_2} + \sum_{k} H^{2n+1}_{2k+1}\vev{\Q'_{2k}\Q_2},
\end{equation}
is a weight $2n+4$ depth $1$ form, we can conclude that all the coefficients $H^{2n+1}_{2k+1}$ are zero and the charge takes the form:
\begin{equation}\label{Even Spin charge}
    \Q_{2n} 
    =  \sum_{j \in \bb{Z} + \lambda} j ^{2n-1} :\chi^-_{-j}\chi^+_{j}:.
\end{equation}
\subsubsection*{Combining the Results}
It is now not difficult to see that the expressions \eqref{Odd Spin Charge} and \eqref{Even Spin charge} can be packaged into one neat formula:
\begin{equation}\label{eq: THE CHARGES}
    \Q_{n-1} 
    = \sum_{j\in \bb{Z}+\lambda} j^{n-2}:\chi^-_{-j} \chi^+_{j}: 
    \;-\; {\bf c}_{n-1}(\lambda) 
    \;,
    \quad\quad 
    \lambda \in\{ 0,\frac{1}{2}\}
    \;.
\end{equation}
In the above expression we have:
\begin{equation}
    {\bf c}_{n-1}(\lambda) = \begin{cases}
         \zeta(1-n)& \lambda =0,\\
         \left(\left(\tfrac{1}{2}\right)^{n-1}-1\right)\zeta(1-n)& \lambda = 1/2.
    \end{cases}
\end{equation}
In the above, we have used the fact that the Riemann zeta function vanishes when evaluated at negative even integers.

\subsubsection{The Way of the Zeta Function}
\label{sec:zeta}

In this section we will obtain the same result as in the previous section, but for arbitrarily twisted representations using zeta-function regularisation.

A Symplectic Fermion representation with arbitrary twist $0<\lambda<1$ has conformal weight $h(\lambda) = \frac{\lambda(\lambda-1)}{2}$. The mode expansions are given by
\begin{equation}
    \chi^{\pm}(z) = \sum_{n \in \bb{Z}\mp \lambda}\chi^{\pm}_nz^{-n-1}.
\end{equation}
Notice that the fields $\chi^+$ and $\chi^-$ have different mode indices now. One obtains a simple irreducible representation of the symplectic fermion algebra, a Fock space as follows:
\begin{equation}
    \mathbb{L}_{\lambda} = \left\{     \chi^-_{-n_1}\dots\chi^-_{-n_\nu}\chi^+_{-m_1}\dots\chi^+_{-m_\mu}\ket{h(\lambda)} \,\,\textbf{;} \,\,\begin{matrix}
        \mu\geq0,\\ \nu\geq0,
    \end{matrix}, \begin{matrix}
        n_i > n_{i+1}>0,\\m_{i}>m_{i+1}>0,
    \end{matrix} \,\, n_i,m_i \in \bb{Z}\pm\lambda\right\}.
\end{equation}
We will take the integral on the cylinder and attempt to regularise it directly. The integral on the cylinder would yield
\begin{equation}
    \Q_{n-1} = \sum_{j\in \bb{Z}-\lambda} j^{n-2}\chi^-_{-j} \chi^+_{j}.
\end{equation}
This expression can be found by looking at \eqref{eq: Bilinear Fields} and noticing that the total derivative terms will not contribute to the integral \eqref{eq: Q = int B}, and then noticing that on the cylinder with coordinate $w$, related to the plane with coordinate $z$ by $z=e^w$, the fields have the following expansion 
\begin{equation}
    \chi^\pm(w) = z \chi^{\pm}(z) = \sum_{n \in \bb{Z}\mp \lambda}\chi^{\pm}_n e^{-wn}.
\end{equation}
By carefully massaging this expression to obtain a normal ordered product we find
\begin{equation}
\begin{split}
\Q_{n-1} &= -(-1)^n \sum_{m \in \bb{Z}^+} (m+\lambda)^{n-1} + 
\sum_{j \in \bb{Z}-\lambda} j^{n-2}:\chi^-_{-j} \chi^+_{j}:\\
&=-(-1)^n\zeta(1-n,\lambda) + \sum_{j \in \bb{Z}-\lambda} j^{n-2}:\chi^-_{-j} \chi^+_{j}:.
\end{split}
\end{equation}
In the above we have noticed that the constant but divergent sum can be formally expressed in terms of the Hurwitz Zeta function. This function is the meromorphic continuation to all $z \neq1$ of the following infinite sum, defined for $\Re(z)>1$
\begin{equation}
\begin{split}
        \zeta(z,a)&= \sum_{n=0}^{\infty}\frac{1}{(n+a)^z} \quad,\quad\Re(z)>1, a\neq0,-1,...
        \;.\\
\end{split}
\end{equation}
It is related to the Riemann zeta function $\zeta$ for the following special values:
\begin{equation}
    \zeta(z,1) = \zeta(z), \quad \zeta(z,\tfrac{1}{2}) = (-1+2^s)\zeta(z)
    \;.
\end{equation} 
For details on the Hurwitz Zeta function, see \S 25.11 of \cite{NIST:DLMF}. A fact that we will use later is that when $z$ is given by the odd integer $1-2n$ and $a$ is given by $1-s$, with $s=0,1/2$, then the Hurwitz zeta function has the following integral representation
\begin{equation}
    \frac{(-1)^n}{2}(2\pi )^n\zeta(1-2n,1-s) = \int_0^{\infty}\frac{u^{2n-1}}{e^{2\pi is+u}-1}du.
\end{equation}
One finds that in a generic twisted sector, that:
\begin{equation}\label{eq: Regularisation Constant}
    {\bf c}_{n-1}(\lambda) = (-1)^n\zeta(1-n,\lambda)= (-1)^n\zeta(1-n,1-\lambda).
\end{equation}
Notice that indeed when $\lambda =0, \frac{1}{2}$ we replicate exactly the result in \eqref{eq: THE CHARGES}.

\section{The Symplectic Fermion GGE}\label{sec:The 
Symplectic Fermion GGE}

The result of the previous section is the existence of a hierarchy of conserved charges given by the zero-modes of quasi-primary fields that are bilinear in the symplectic fermion fields. These charges are denoted by $\Q_{n}$ and are defined in \eqref{eq: THE CHARGES}.  When the subscript $n=2k$ is even, the commutation relations with the symplectic fermion modes are
\begin{equation}\label{eq: Even Commuter}
    [\Q_{2k},\chi^{\pm}_n] = \mp n^{2k}\chi^{\pm}_n,
\end{equation}
while for odd subscripts $n=2k-1$ the relations are
\begin{equation}\label{eq: Odd Commuter}
    [\Q_{2k-1},\chi^{\pm}_n] = -n^{2k-1}\chi^{\pm}_n.
\end{equation}
We now construct the extended characters by inserting the first $N$ charges into the trace, each accompanied by its own chemical potential
\begin{equation}
    \Psi_{\lambda,s}(\t,\mu_0,\mu_2,\ldots,\mu_N)
    := \Tr_{\lambda,s} \!\left(
    q^{L_0-\frac{c}{24}} e^{\sum_{k\geq2}\mu_k \Q_k}
    \right),
    \quad \lambda,s = 0,\tfrac{1}{2}.
\end{equation}
Using the commutation relations above, we obtain exact expressions for the GGEs 
\begin{equation}\label{eq: GGE general}
    \Psi_{\lambda,s}(\boldsymbol{\mu})
    = e^{-\sum_{n =1}^N\mu_n{\bf c}_{n}(\lambda)}
    \prod_{n\in \bb{N-\lambda}}
    \left(1-e^{2\pi is} e^{\sum_{k=1}^{N}\mu_k n^k}\right)
    \left(1-e^{2\pi is} e^{-\sum_{k=1}^{N}(-1)^k\mu_k n^k}\right),
\end{equation}
where, for conciseness, we have defined $\mu_1 = 2\pi i \t$. 

It is also important to discuss convergence. We will always assume that there are only a finite number of non-zero chemical potentials, and further that at least one non-zero odd spin charge in the GGE.
If the highest-spin charge inserted is odd, 
then convergence requires
\begin{align}
    \Re(\mu_{2K-1})<0
    \;.
    \label{eq:condition1}
    \end{align}
If, in addition, there is a finite set of higher even-spin charges $\Q_{2n_i}$ with $2n_i > 2K-1$, we further require
\begin{equation}
    \Re(\mu_{2n_i})=0\;,
\label{eq:condition2}
    \end{equation}
so that the GGE is convergent by the terms in it being exponentially suppressed. 

We will derive the exact modular transformation of this GGE in the next section and also analyse its asymptotic behaviour.

\subsection{Exact Transformation of Symplectic Fermion GGE}\label{sec: Exact Transformation of Symplectic Fermion GGE}
In this section we will obtain an exact transformation of the GGE. We highlight in advance the final result \eqref{eq: Final Transformation} and \eqref{eq:roots}. This proof follows those in \cite{Downing:2023lop,zagier:hal-03494849}, but with some important differences in the details.

It will be helpful to introduce constants $\alpha_j$ and parametrise the 
chemical potentials by
\begin{equation}
    \mu_k = 2\pi i \alpha_k \t^k
    \;.
\end{equation}
 For generality, we will leave $\a_1$ free, but later, when we perform an asymptotic analysis of the transformation, we will set $\a_1 = 1$. 
 For the sake of compactness, we introduce the following notation: 
\begin{equation}
\prod_{\pm}S_{\pm} = S_+S_- \;,\quad\sum_{\pm}S_{\pm} = S_++S_-\;.
\end{equation}
In terms of these, the GGE \eqref{eq: GGE general} becomes
\begin{equation}\label{eq: GGE alphas.}
    \Psi_{\lambda,s}(\tau,\boldsymbol{\alpha})= e^{-2\pi i\sum_{j =1}^N\t^j\a_j{\bf c}_{j}(\lambda)}\prod_{n\in \bb{N-\lambda}>0}\prod_{\pm} \left(1-e^{2\pi is} e^{\pm2\pi i\sum_{j=1}^{N}(\pm1)^j\t^j\a_jn^j}\right).
\end{equation}

For reasons that will become clear in the subsequent paragraphs, let us introduce a real, positive parameter $\theta$ into the GGE, (and suppress for the moment the dependence on $\t$)
\begin{equation}\label{eq: GGE with theta and alpha}
    \Psi_{\lambda,s}(\theta|\boldsymbol{\alpha})= e^{-2\pi i\sum_{j =1}^N\t^j\a_j{\bf c}_{j}(\lambda)}\prod_{n\in \bb{N-\lambda}>0}\prod_{\pm} \left(1-e^{2\pi is} e^{\pm2\pi i\theta\sum_{j=1}^{N}(\pm1)^j\t^j\a_jn^j}\right).
\end{equation}
We can recover our original GGE simply by setting $\theta=1$. The convergence conditions
\eqref{eq:condition1} and \eqref{eq:condition2}
become, where $K$ is the largest value such that $\alpha_{2K-1}$ is non-zero, 
\begin{equation}\label{eq: Expanded domain}
\begin{split}
    &\Im(\a_{2K-1}\t^{2K-1})>0\;,\\
    &\Im(\a_{2n_i}\t^{2n_i})=0\;, \quad 2n_i>2K-1.
\end{split}
\end{equation}
However, in this calculation we will work in a more restricted domain, where all the chemical potentials are constrained. 
We always require there to be only a finite set of non-zero $\mu_n$. If $K$ is the largest value such that $\alpha_{2K-1}$ is non-zero, 
then we require
\begin{equation}\label{eq: Restricted Domain}
\begin{split}
&\Im(
\,\a^{\vphantom{\phi}}_{2K-1}\t^{2K-1})>0,  \\
&\Im(
\,\a_{2n-1}\t^{2n-1})\geq0, \quad \forall\,\,\, n<K,\\
&\Im(
\,\a_{2n}\t^{2n})=0, \quad \forall \,\,\,n.\\
\end{split}
\end{equation}
At the end our result still makes sense for the larger domain \eqref{eq: Expanded domain} and we can argue that by analytic continuation it is correct for that larger domain.

We take the log of the expression \eqref{eq: GGE with theta and alpha}, to obtain  a sum over logs and a sum of Hurwitz zeta functions. We will denote by $\CB_{\lambda,s}$ the sum over the logs.
\begin{equation}\label{eq: Definition of SCB}
\begin{split}
    \CB_{\lambda,s} &= \log(\Psi_{\lambda,s}(\theta|\boldsymbol{\alpha})) +2\pi i\sum_{j=1}^{K}\t^{2j-1}\a_{2j-1}\zeta(1-2j,1-\lambda)\\
    &=\sum_{n \in \bb{Z}-\lambda>0}\sum_{\pm}\log(1-e^{2\pi is}e^{\pm2\pi i \theta\sum_{j=1}^N(\pm1)^j\t^j \a_jn^j}).
\end{split}
\end{equation}
As before, $2K-1$ is the spin of the highest odd-spin charge in the GGE. Here we will pick the branch of $\log$ such that $\log(-1) = -\pi i$, although it will not matter in the end since we will exponentiate our answer. To apply the Poisson resummation formula, we need to rewrite this as a sum over the entire integer lattice, and not just $\bb{Z}-\lambda>0$. Formally, we can use
\begin{align}
    \log(1 + e^x) =
    \tfrac 12 \log(1+e^x) 
    + \tfrac 12 \log(1 + e^{-x}) + \tfrac 12 \log x\;,
\end{align}
to rewrite 
$\CB_{\lambda,s}$ (where $\lambda\in\{0,1/2\}$)  as
\begin{equation}
\begin{split}
\CB_{\lambda,s} &= \frac{1}{2}{\sideset{}{'}\sum_{n \in \bb{Z}-\lambda}}\sum_{\pm}\log(1-e^{2\pi is}e^{\pm2\pi i \theta\sum_{k=1}^N(\pm1)^j\t^j \a_jn^j})\\
&+\delta_{\lambda,0}\delta_{s,\frac{1}{2}}\log(2) + \pi i\sum_{n\in \bb{Z}-\lambda<0} \left[2 s - 1 +\theta \sum_{j=1}^N(1-(-1)^j)\t^j\a_jn^j\right].
\end{split}
\end{equation}
The prime indicates that the case $n=0, s=0$ is excluded. However, the sums in the first and second lines both diverge as $n\to-\infty$.
It is here where the introduction of the parameter $\theta$ becomes useful, as we can differentiate twice with respect to it to do away with the divergent piece. First, we take two derivatives of $\CB_{\lambda,s}$ in \eqref{eq: Definition of SCB}, and only then rewrite it as a sum over the lattice which gives
\begin{equation}
\del^2_{\theta}\CB_{\lambda,s} = \frac{1}{2}\sideset{}{'}\sum_{n \in\bb{Z}-\lambda}\sum_{\pm}\del^2_\theta \log(1-e^{2\pi i s}e^{\pm2\pi i \theta\sum_{j=1}^N(\pm1)^j\t^j\a_jn^j})
\end{equation}
We then introduce the functions $p_{\theta,s}^\pm$ and $F_{\theta,s}^\pm$,
\begin{equation}
p_{\theta,s}^{\pm}(t) = 1- e^{2\pi i s}e^{\pm2\pi i \theta\sum_{j=1}^N(\pm1)^j(it)^j\a_j} 
\;,
\quad\quad F_{\theta,s}^{\pm}(t)= \del^2_{\theta} \log(p^{\pm}_{\theta,s}(t))
\;,
\end{equation}
where the variable $t$ is related to $\t$ by $\t = it$. We can rewrite the infinite series as
\begin{equation}\label{eq: SCA = sum log p = sum F}
\CB_{\lambda,s}
=  
\frac 12
\sideset{}{'}\sum_{n \in \bb{Z}-\lambda} \sum_{\pm}\del^2_{\theta}\log(p_{\theta,s}^\pm(nt))
= 
\frac 12
\sideset{}{'}\sum_{n \in \bb{Z}-\lambda} \sum_{\pm}F_{\theta,s}^{\pm}(nt).
\end{equation}
The above series converges in the domain \eqref{eq: Expanded domain} with $\t = it$. The Poisson resummation formula for a function $f(\nu t)$ on the integer shifted lattice is
\begin{equation}
    \sum_{\nu \in \bb{Z}+\lambda}f(\nu t) = \frac{1}{t}\sum_{m \in \bb{Z}} \tilde{f}\left(\frac{m}{t}\right)e^{-2\pi i m\lambda}\;,
    \label{eq:poisson}
\end{equation}
where the tilde denotes the Fourier transformation
\begin{equation}
    \tilde{f}(y)= \int_{-\infty}^{\infty}f(t)\,e^{2\pi i ty}\,dt
    \;.
\end{equation}
To use \eqref{eq:poisson} in the case $\lambda=s=0$, we have to sum over the whole integer lattice with $f(x)$ a smooth function, so we add in the limiting term as $n\to 0$ of $F^\pm_{\theta,0}(nt)$, and  define
\begin{align}
    F^\pm_{\theta,0}(0) = 
    \lim_{x\to 0} F^\pm_{\theta,0}(x)
    =-\frac{1}{\theta^2}
\end{align}
and now we consider
\begin{equation}\label{eq: A = 2 del SCB - 2 theta squared}
\CA_{\lambda,s} 
= 2 \del^2_{\theta}\CB_{\lambda,s}-\frac{2}{\theta^2}\delta_{\lambda,0}\delta_{s,0}\\
=\sum_{n \in \bb{Z}-\lambda} \sum_{\pm}F_{\theta,s}^{\pm}(nt)
\;.
\end{equation}

Based on \eqref{eq: SCA = sum log p = sum F}, we need to calculate the fourier transforms $\tilde{F}^{\pm}_{\theta,s}(y)$ in the regimes $y<0$, $y=0$ and $y>0$, which we do separately now.
\subsection*{\boldmath $\bullet\quad y \neq 0$}
We integrate by parts to obtain
\begin{equation}
\label{eq:tildeF}
    \tilde{F}^{\pm}_{\theta,s}(y)
    =
    \int_{-\infty}^{\infty}\del^2_{\theta}\log(p^\pm_{\theta,s}(t))\,e^{2\pi i t y}\,dt
    =
    -\frac{1}{2\pi i y}\int_{-\infty}^{\infty}\del^2_{\theta}\left(\frac{\dot{p}^{\pm}_{\theta,s}(t)}{p^{\pm}_{\theta,s}(t)}\right)\,e^{2\pi i t y}\,dt,
\end{equation}
where the dot on $\dot p^{\pm}_{\theta,s}$ denotes a $t$ derivative. Depending on whether $y>0$ or $y<0$, we will extend the integral into the upper or lower half plane so that the integral on the contour going out to infinity is exponentially suppressed by $e^{2\pi i ty}$.
\subsubsection*{\boldmath $\bullet\quad y>0$} 
We can close the contour in the upper half plane and the integral is given by a sum over residues at poles in the upper half plane. By inspection, poles can only come from zeroes of $p^\pm_{\theta,s}(t)$, and around each pole the $\theta$-derivatives can be pulled outside the integral making the result immediate: if $p^\pm_{\theta,s}(t)$ has a zero at $t=t_0^\pm$ of order $M$, then the contribution to the integral \eqref{eq:tildeF} from the residue at $t_0^\pm$, coming from an anti-clockwise contour is
\begin{equation}
    \frac{-M}{y}\del^2_{\theta}\left(e^{2\pi i t_0^{\pm}y}\right)
    \;.
\end{equation}
The condition for 
$t_0^{\pm}$ to be a zero of $p^\pm_{\theta,s}$ is 
\begin{equation}
    \pm\sum_{j=1}^N (\pm1)^j\a_j (it_0^{\pm})^j = \frac{n}{\theta}, \quad n \in \bb{Z}+s.
\end{equation}
Rather than trying to keep track of the order of any particular zero, we will simply sum (with multiplicity) and obtain
\begin{equation}\label{eq: F tilde y geq 0}
    \tilde{F}^{\pm}_{\theta,s}(y) = - \sum_{n\in \bb{Z}+s} \sum_{\tilde{\om}^\pm_l(n/\theta)\in \h^+}\frac{1}{y}\del^2_{\theta}e^{2\pi i \tilde{\om}_l^{\pm}(n/\theta)y},
\end{equation}
where $\tilde{\om}_j^{\pm}(\kappa)$ satisfies
\begin{equation}\label{eq: 14}
    \om= \tilde{\om}_l^{\pm}(\kappa):\quad \pm\sum_{j=1}^N (\pm1)^j\a_j (i\om)^j = \kappa
\end{equation}
Here $l$ labels the roots of the equations, $l=0,...,N-1$, and we also define
\begin{equation}
    \bb{H}^{+} = \{z\in \bb{C}| \Im(z)>0\}\quad,\quad\bb{H}^{-} = \{z\in \bb{C}| \Im(z)<0\}.
\end{equation}
Let us emphasise, that by the sum over the roots we mean
\begin{equation}
    \sum_{\tilde{\om}^\pm_l(n/\theta)\in \h^+} = \sum_{\pm}\sum_{\substack{l=0\\ \tilde \omega^\pm_l(n/\theta)\in \h^+}}^{N-1},
\end{equation}
that is, this notation is a shorthand for a sum over all the roots that lie in the upper half plane of both associated polynomials. We'll use a similar expression for $\h^-$. 
\subsubsection*{\boldmath $\bullet\quad y<0$}
For the case of $y<0$, one performs similar steps in the lower half plane with clockwise contours to obtain instead
\begin{equation}\label{eq: F tilde y leq 0}
    \tilde{F}^{\pm}_{\theta,s}(y) =  \sum_{n\in \bb{Z}+s} \sum_{\tilde{\om}_l(n/\theta)\in \h^-}\frac{1}{y}\del^2_{\theta}e^{2\pi i \tilde{\om}_l^{\pm}(n/\theta)y},
\end{equation}
\subsubsection*{\boldmath $\bullet\quad y=0$}
Here the exponential in the integrand $e^{2\pi i t y}=1$, so there is no need to extend the integration contour into the complex plane. We will leave the expression as is:
\begin{equation}\label{eq: F tilde y eq 0}
    \tilde{F}^{\pm}_{\theta,s}(0) = \int_{\bb{R}} \del^2_{\theta}\log(p^{\pm}_{\theta,s}(u))du.
\end{equation}

Let us bring the equations \eqref{eq: F tilde y eq 0}, \eqref{eq: F tilde y geq 0}, \eqref{eq: F tilde y leq 0} and \eqref{eq: SCA = sum log p = sum F}, together to obtain the intermediate result:
\begin{equation}\label{eq: Intermediate Result}
\begin{split}
\CA_{\lambda,s}(\t) 
&=\frac{1}{t}\int_{\bb{R}}\del^2_{\theta}\sum_{\pm}\log(p^\pm_{\theta,s}(u))du\\
&\quad-\sum_{m=1}^{\infty}\frac{1}{m}\del^2_{\theta}\left(\sum_{n \in \bb{Z}-s}\sum_{\tilde{\om}_l^\pm(n/\theta)\in \h^-}e^{-2\pi i  \tilde{\om}_l^{+}(n/\theta) m/t}+e^{-2\pi i  \tilde{\om}_l^{-}(n/\theta) m/t}\right)e^{2\pi i  m \lambda}\\
&\quad-\sum_{m=1}^{\infty}\frac{1}{m}\del^2_{\theta}\left(\sum_{n \in \bb{Z}-s}\sum_{\tilde{\om}_l^\pm(n/\theta)\in \h^+}e^{2\pi i  \tilde{\om}_l^{+}(n/\theta) m/t}+e^{2\pi i  \tilde{\om}_l^{-}(n/\theta) m/t}\right)e^{-2\pi i m \lambda}.\\
\end{split}
\end{equation}
We have used the fact that $\lambda \in\{0,1/2\}$ to allow us to
swap $\lambda$ with $-\lambda$ in the above expression when swapping 
the sum over $m<0$  into a sum over $m>0$. Since the roots $\tilde{\om}_l^{\pm}(n/\theta)$ lie in the upper/lower half-planes, each sum will converge exponentially. This is useful as it means we can swap the $m$ and $n$ summations, and evaluate the $m$ sum by recognising it as the Taylor series for $\log(1-x)$.
\begin{equation}
\begin{split}
\CA_{\lambda,s}(\t) 
&=\frac{1}{t}\int_{\bb{R}}\del^2_{\theta}\left(\log(p^+_{\theta,s}(u))+\log(p^-_{\theta,s}(u))\right) du\\
&\quad+\left(\sum_{n \in \bb{Z}-s}\sum_{\tilde{\om}_l^\pm(n/\theta)\in \h^-}\del^2_{\theta}\log (1-e^{2\pi i   \lambda}e^{-2\pi i  \tilde{\om}_l^{+}(n/\theta) /t})+\del^2_{\theta}\log(1-e^{2\pi i   \lambda}e^{-2\pi i  \tilde{\om}_l^{-}(n/\theta) /t})\right)\\
&\quad+\left(\sum_{n \in \bb{Z}-s}\sum_{\tilde{\om}_l^\pm(n/\theta)\in \h^+}\del^2_{\theta}\log(1-e^{2\pi i   \lambda}e^{2\pi i  \tilde{\om}_l^{+}(n/\theta) /t})+\del^2_{\theta}\log(1-e^{2\pi i   \lambda}e^{2\pi i  \tilde{\om}_l^{-}(n/\theta) /t})\right).\\
\end{split}
\end{equation}
Once again, we used $\lambda \in\{0,1/2\}$ to simplify the exponential that involves $\lambda$. Now consider \eqref{eq: 14}, and define
\begin{equation}
    \om_l^\pm(\kappa):=\frac{1}{t}\tilde{\om}_l^\pm(\kappa) = -\frac{1}{i\t}\tilde{\om}_l^\pm(\kappa).
\end{equation}
Then the roots $\om_l^\pm(\kappa)$ satisfy
\begin{equation}\label{eq: Polynomial ting}
    \om = \om_l^\pm(\kappa): \quad \pm\sum_{j=1}^N(\pm1)^j\a_j(\t \om)^j= \kappa.
\end{equation}
Notice also that if $\om_l^+(\kappa)$ satisfies the $(+)$ equation, then we can define $\om = -\om_l^+(-\kappa)$ which solves the $(-)$ equation. As such, we will pair each root of each equation in that fashion
\begin{equation}\label{eq: Pairing up roots}
    \om_l^{\pm}(\kappa) = - \om^{\mp}_l(-\kappa).
\end{equation}
We collect all the relevant properties of these polynomial roots in Appendix \ref{app: Properties of Roots}. Given this fact, we can rewrite our sum in terms of the roots $\om_l^\pm(\kappa)$, and have all roots in the upper half plane. One can also perform the substitution $x = iu/\t$ to simplify the integral
\begin{equation}\label{eq: Done in Derivative space}
\begin{split}
\CA_{\lambda,s}(\t) 
&=2\int_0^\infty\sum_{\pm}\del^2_{\theta}\log(1-e^{2\pi is}e^{\pm2\pi i \theta\sum_{j=1}^N(\pm1)^j\t^j\a_jx^j})dx\\\
&\quad+2\sum_{n \in \bb{Z}-s}\sum_{\om_l^\pm(n/\theta)\in \h^+}\sum_\pm\del^2_{\theta}\log(1-e^{2\pi i   \lambda}e^{2\pi i  {\om}_l^{\pm}(n/\theta) })\,,\\
\end{split}
\end{equation}
where we noticed that the integrand in the first line is an even function of $x$ to rewrite the integral as only over the positive reals. Now this integral converges if we were to remove the $\del^2_\theta$, so we can safely pull it out of the integral. Furthermore, given the asymptotic behaviour of the roots $\om_l^{\pm}$, outlined in Appendix \ref{app: Properties of Roots}, and the fact that we only choose roots in the upper half plane, then it is the case that the sum in the second line also converges with or without the derivatives, due to the exponential suppression of the terms at large $n$, and so we can safely pull the derivatives out. Now recall \eqref{eq: A = 2 del SCB - 2 theta squared}, which tells us that
\begin{equation}
\del^2_{\theta}\CB_{\lambda,s}=\frac{1}{2}\CA_{\lambda,s} + \frac{1}{\theta^2}\delta_{\lambda,0}\delta_{s,0}.
\end{equation}
Since we can pull out $\theta$ derivatives without spoiling convergence, we can integrate the above expression with respect to $\theta$ so as to find
\begin{equation}\label{eq: B + log = int + roots + consts}
\begin{split}
& \CB_{\lambda,s} +\log(\theta)\delta_{\lambda,0}\delta_{s,0} \\ &=\int_0^\infty\sum_{\pm}\log(1-e^{2\pi is}e^{\pm2\pi i \theta\sum_{j=1}^N(\pm1)^j\t^j\a_jx^j})dx \\
&\quad+\sum_{n \in \bb{Z}-s}\sum_{\om_l^\pm(n/\theta)\in \h^+}\log(1-e^{2\pi i   \lambda}e^{2\pi i  {\om}_l^{+}(n/\theta) })+\log(1-e^{2\pi i   \lambda}e^{2\pi i  {\om}_l^{-}(n/\theta) })\\
&\quad+ \CC_{\lambda,s}\theta + \CD_{\lambda,s},
\end{split}
\end{equation}
where $\CC_{\lambda,s}$ and $\CD_{\lambda,s}$ are integration constants. We will determine these constants by performing an asymptotic expansion around $\theta \to \infty$. For compactness, let us denote the integral by $\CI_{\lambda,s}$ and the sum over roots to be $\CR_{\lambda,s}$. We need to analyse now the equation
\begin{equation}\label{eq: The equation we're expanding}
    \CB_{\lambda,s} +\log(\theta)\delta_{\lambda,0}\delta_{s,0} = \CI_{\lambda,s} +\CR_{\lambda,s} + \CC_{\lambda,s}\theta + \CD_{\lambda,s}.
\end{equation}
We will look at each piece in turn.
\subsection*{\boldmath $\bullet\quad\CB_{\lambda,s}$}
Recall from \eqref{eq: Definition of SCB} that
\begin{equation}
    \CB_{\lambda,s}=\sum_{n \in \bb{Z}-\lambda>0}\sum_{\pm}\log(1-e^{2\pi is}e^{\pm2\pi i \theta\sum_{j=1}^N(\pm1)^j\t^j \a_jn^j}).
\end{equation}
Within the domain \eqref{eq: Restricted Domain} we see that the sum has exponential suppression in each term as $\theta \to \infty$, and as such will not contribute to the integration constants.
\begin{equation}\label{eq: Expansion of SCB Final}
    \CB_{\lambda,s}\sim 0.
\end{equation}
\subsection*{\boldmath $\bullet\quad\CI_{\lambda,s}$}
The integral is
\begin{equation}\label{eq: SCI = integral}
    \CI_{\lambda,s} = \int_0^\infty\sum_{\pm}\log(1-e^{2\pi is}e^{\pm2\pi i \theta\sum_{j=1}^N(\pm1)^j\t^j\a_jx^j})dx. 
\end{equation}
Recall that the $l=0,1,...,N-1$ roots $\om^\pm_l(\kappa)$ are solutions to
\begin{equation}
    \om = \om^\pm_l(\kappa): \quad \pm \sum_{j=1}^N(\pm1)^j\t^j\a_j\om^j=\kappa.
\end{equation}
Denote the $l=0$ root, $\om^{\pm}_0(\kappa)$, to be the distinguished root that is continuously connected to $0$ when $\kappa = 0$. In the $(+)$ term in \eqref{eq: SCI = integral}, we will perform the substitution $x = \om_0^+(\frac{iu}{2\pi \theta})$, and in the $(-)$ term in \eqref{eq: SCI = integral} we will perform the substitution $x = \om_0^-(\frac{iu}{2\pi \theta})$. We find then that
\begin{equation}
\int_{0}^{\infty} \log\bigl(1-e^{2\pi i s}e^{-u}\bigr)\frac{d}{du}\left(\om_0^{+}\!\left(\tfrac{iu}{2\pi \theta}\right)+\om_0^{-}\!\left(\tfrac{iu}{2\pi \theta}\right)\right)  du.
\end{equation}
In the above we deformed the contour of the integral to the positive real axis. We can do this because the only poles in the integrand lie on the imaginary axis and since the roots are continuous; given the assumptions about our domain \eqref{eq: Restricted Domain}, we can show that when $x>0$ then $\Re(u)>0$, so $u$ stays on the right hand side of the plane and goes out to infinity and does not ever cross the imaginary axis. Performing integration by parts we obtain
\begin{equation}
    - \int_0^{\infty}\frac{1}{e^{2\pi is}e^{u}-1}(\om_0^{+}\!\left(\tfrac{iu}{2\pi \theta}\right)+\om_0^{-}\!\left(\tfrac{iu}{2\pi \theta}\right))du.
\end{equation}
The sum of $\om_0^{\pm}$ is an odd function, a fact apparent from \eqref{eq: Pairing up roots}, so we will suppose it has a power series expansion like
\begin{equation}
    \om_0^{+}(x) + \om_0^{-}(x) = \sum_{n=1}^{\infty}O_{2n-1}x^{2n-1},
\end{equation}
which may have a finite radius of convergence, so when we use this expansion to swap the integral and sum, we will in general have an asymptotic result. The integral expansion is
\begin{equation}
    \sum_{n=1}^{\infty}O_{2n-1}\left(\frac{i(-1)^n}{(2\pi \theta)^{2n-1}}\right)\int_0^{\infty}\frac{u^{2n-1}}{e^{2\pi is+u}-1} = \pi i \sum_{n=1}^{\infty}O_{2n-1}\zeta(1-2n,1-s)\theta^{1-2n}.
\end{equation}
Here we have used the definition of the Hurwitz Zeta function
\begin{equation}
\int_0^{\infty}\frac{u^{2n-1}}{e^{2\pi is+u}-1}du = \frac{(-1)^n}{2}(2\pi)^{2n}\zeta(1-2n,1-s).
\end{equation}
Hence, the asymptotic expansion of the integral expression is
\begin{equation}\label{eq: Expansion of SCI Final}
\begin{split}
\CI_{\lambda,s}\sim \pi i \sum_{n=1}^{\infty}O_{2n-1}\zeta(1-2n,1-s)\theta^{1-2n}
\end{split}
\end{equation}
\subsection*{\boldmath $\bullet\quad\CR_{\lambda,s}$}
We will need in this section some facts from Appendix B of \cite{Downing:2023lop}, which comes from \cite{zagier2006mellin}.\\
\textbf{Fact 1:} If we start with a function $f(t)$ which has asymptotic expansion
\begin{equation}
    f(t)\sim \sum_{n=0}^\infty b_n t^n, \quad t\to0,
\end{equation}
then it is true that
\begin{equation}
    \sum_{m=0}^{\infty}f((m+a)t) \sim \frac{1}{t}\int_0^{\infty}f(x)dx + \sum_{n=0}^{\infty}\zeta(-n,a)b_nt^n\quad,\quad t\to 0^+.
\end{equation}
Since we have sums over $\bb{Z}-s$ with $s = 0,\frac{1}{2}$, we need only the result with $a = \frac{1}{2},1$. That is
\begin{equation}\label{eq: Basic asymptotic, no log}
    \sum_{m=0}^{\infty}f((m+a)t) \sim \frac{1}{t}\int_0^{\infty}f(x)dx-\frac{1}{2}b_0\delta_{a,1} + \sum_{n=1}^{\infty}\zeta(1-2n,a)b_{2n-1}t^{2n-1},\quad t\to 0^+.
\end{equation}
\textbf{Fact 2:} If we start with a function $f(t)$ which has asymptotic expansion including a log term
\begin{equation}
    f(t) \sim b \log(t) + \sum_{n=0}^{\infty}b_nt^n \quad,\quad t\to0
\end{equation}
then as $t\to 0^+$, and $a=\frac{1}{2},1$,
\begin{equation}\label{eq: asymptotic log}
    \sum_{m=0}^{\infty}f((m+a)t) \sim \frac{1}{t}\int_0^{\infty}f(x)dx+\frac{1}{2}\delta_{a,1}(b\log(2\pi/t)-b_0) +\frac{b}{2}\delta_{a,\frac{1}{2}}\log(2)+ \sum_{n=1}^{\infty}\zeta(1-2n,a)b_{2n-1}t^{2n-1}.
\end{equation}
Recall that the sum over roots in \eqref{eq: B + log = int + roots + consts}, is
\begin{equation}
    \CR_{\lambda,s} := \sum_{n \in \bb{Z}-s}\sum_{\om_l^\pm(n/\theta)\in \h^+}\log(1-e^{2\pi i   \lambda}e^{2\pi i  {\om}_l^{+}(n/\theta) })+\log(1-e^{2\pi i   \lambda}e^{2\pi i  {\om}_l^{-}(n/\theta) }).
\end{equation}
We see that based on fact 1 and fact 2, that it will be easier to set $\Theta = \theta^{-1}$ and expand around $\Theta \to 0$. Let us denote the contributions to $\CR_{\lambda,s}$ that come from the roots $\om_0^{\pm}(\kappa)$, that are continuously connected to $0$ when $\kappa = 0$, by $\CR^{(0)}_{\lambda,s}$. Let us also define new functions
\begin{equation}
\begin{split}
g_l^\pm(t) &= \log(1-e^{2\pi i \lambda}e^{\pm2\pi i \om_l^+(t)}) + \log(1-e^{2\pi i \lambda}e^{\mp 2\pi i \om_l^{-}(t)}),\\
g_l^0 &= \log(1-e^{2\pi i\lambda}e^{2\pi i \om_l^+(0)}) + \log(1-e^{2\pi i\lambda}e^{2\pi i \om_l^-(0)}).
\end{split}
\end{equation}
Given that for $\kappa>0$ and for $l\neq0$, $\Im(\om_l^+(\kappa))>0$ then $\Im(\om_l^-(\kappa))<0$ and vice versa, we can show that:
\begin{equation}
\CR_{\lambda,s} = \CR_{\lambda,s}^{(0)} + \delta_{s,0}\sum_{l \neq 0}\sum_{\om_l^\pm(0)\in \h^+} g_l^0+ \sum_{n \in \bb{Z}-s>0}\sum_{\pm}\sum_{\om_l^\pm(n\Theta)\in \h^+, l \neq 0}g_l^\pm(n\Theta).
\end{equation}
Let us focus on the piece not associated to the root $l=0$. What we mean by this expression is that for each $n$, and each $l\neq0$, we have $g_l^+$ contribute if $\omega_l^+(n\Theta)\in \h^+$ and $\omega_l^-(n\Theta)\in \h^-$ or $g_l^-$ contribute if $\omega_l^+(n\Theta)\in \h^-$ and $\omega_l^-(n\Theta)\in \h^+$. Since both of $g_l^\pm(t)$ are even functions, the expansion of their sum is given by 
\begin{equation}
    2g_l^0 +\sum_{n=1}^\infty b_{2n}t^{2n},
\end{equation}
so by fact 1 \eqref{eq: Basic asymptotic, no log}, we find that 
\begin{equation}
    \CR_{\lambda,s} \sim \CR_{\lambda,s}^{(0)} +\frac{1}{\Theta} \int_0^\infty \sum_{l\neq0}\sum_\pm g_l^\pm(x)dx.
\end{equation}
Let us focus on the $(+)$ integral. We have that
\begin{equation}
    \int_0^\infty g_l^+(x)dx =  \int_0^\infty\log(1-e^{2\pi i \lambda}e^{2\pi i \om_l^+(x)}) + \log(1-e^{2\pi i \lambda}e^{-2\pi i \om_l^{-}(x)})dx.
\end{equation}
Suppose $\Im(\om_l^+(x))>0$, and by appendix \ref{app: Properties of Roots} that $\Im(\om_l^-(x))<0$. In the first integral perform the substitution $\om_l^+(x) = iw$ and in the next $\om_l^-(x) = -iw$.  If instead we had $\Im(\om_l^+(x))<0$ we would perform the substitutions the other way around, but the subsequent steps would be the same. Integration by parts yields
\begin{equation}
\int_0^{\infty}g_l^+(x)dx = -2\pi \int_{-i\om_l^+(0)}^{-i\om_l^+(\infty)}\frac{x(iw)}{e^{2\pi i\lambda}e^{2\pi w}-1}dw+2\pi \int_{i\om_l^-(0)=-i\om_l^+(0)}^{i\om_l^-(\infty)}\frac{x(iw)}{e^{2\pi i\lambda}e^{2\pi w}-1}dw.
\end{equation}
In the above we used that $x(-iw) = - x(iw) $, a fact apparent from \eqref{eq: Polynomial ting}. Considering Appendix \ref{app: Properties of Roots}, which as in the previous paragraph tells us that $\om_l^\pm(x)$ will lie on the upper or lower half the plane, we know that the integral will never hit the imaginary axis where the poles of the integrand lie, and also that the integration contours stay firmly on the right hand side of the complex $w$ plane, we can argue that the integration contours can be deformed onto each other, and exactly cancel one another. Similar reasoning shows that the $(-)$ integral vanishes too. 
\begin{equation}
    \int_0^\infty g_l^\pm(x)dx =0.
\end{equation}
Now we need only deal with the roots continuously connected to zero.
\begin{equation}
    \CR_{\lambda,s} \sim \CR_{\lambda,s}^{(0)}.
\end{equation}
We know that
\begin{equation}
    \CR_{\lambda,s}^{(0)} = \sum_{n \in \bb{Z}-s}\sum_{z_0^{\pm}(n\Theta)\in \bb{H}^+}\log(1-e^{2\pi i   \lambda}e^{2\pi i  {\om}_0^{+}(n\Theta) })+\log(1-e^{2\pi i   \lambda}e^{2\pi i  {\om}_0^{-}(n\Theta) }).
\end{equation}
We can rewrite this as a half-lattice sum
\begin{equation}
\begin{split}
 \CR_{\lambda,s}^{(0)}  &= \sum_{n \in \bb{Z}-s>0}\sum_{\om_0^\pm(n\Theta)\in \h^+} \log(1-e^{2\pi i\lambda}e^{2\pi i \om_0^+(n\Theta)})+\log(1-e^{2\pi i\lambda}e^{2\pi i \om_0^-(n\Theta)})\\
 &+\sum_{n \in \bb{Z}-s>0}\sum_{\om_0^\pm(n\Theta)\in \h^-} \log(1-e^{2\pi i\lambda}e^{-2\pi i \om_0^-(n\Theta)})+\log(1-e^{2\pi i\lambda}e^{-2\pi i \om_0^+(n\Theta)})\\
\end{split}
\end{equation}
Appendix \ref{app: Properties of Roots} tells us that the roots $\om_0^\pm(\kappa)$ always lie in the lower half plane for $\kappa>0$, so we just have:
\begin{equation}
\begin{split}
 \CR_{\lambda,s}^{(0)}  
 =\sum_{n \in \bb{Z}-s>0}\log(1-e^{2\pi i\lambda}e^{-2\pi i \om_0^-(n\Theta)})+\log(1-e^{2\pi i\lambda}e^{-2\pi i \om_0^+(n\Theta)})\\
\end{split}
\end{equation}
Consider the function
\begin{equation}
    G_{\lambda}(t) = \log(1-e^{2\pi i\lambda }e^{-2\pi i \om_0^{-}(t)}) +\log(1-e^{2\pi i\lambda }e^{-2\pi i \om_0^{+}(t)}).
\end{equation}
It is not difficult to show that $G_\lambda(t)$ can be written as
\begin{equation}
    G_\lambda(t) = \CE^\lambda(t)+2\log(2) + \delta_{\lambda,0}\log\left(\frac{\pi i t}{\a_1 \t}\right) - \pi i (\om_0^{+}(t)+\om_0^{-}(t)),
\end{equation}
where $\CE^\lambda(t)$ is the even function which vanishes as $t \to 0$ and is given by
\begin{equation}
    \pi  i (\om_0^{+}(t)+\om_0^{-}(t)) + \log\left(\frac{1}{2}\left(\frac{\a_1 \t}{\pi i t}\right)^{\delta_{\lambda,0}}(1-e^{2\pi i \om_0^{+}(t)})\right)+\log\left(\frac{1}{2}\left(\frac{\a_1 \t}{\pi i t}\right)^{\delta_{\lambda,0}}(1-e^{2\pi i \om_0^{-}(t)})\right).
\end{equation}
From the above we deduce that $G_\lambda$ has an asymptotic expansion of the form
\begin{equation}
G_\lambda(t) \sim 2\log(2) + 2\delta_{\lambda,0}\log\left(\frac{\pi i t}{\a_1 \t}\right)-\sum_{n=1}^{\infty}O_{2n-1}t^{2n-1} + \sum_{n=1}^{\infty}\CE^\lambda_{2n}t^{2n},
\end{equation}
where $\CE_{2n}^\lambda$ are the coefficients in the expansion of the even function. We also used the definition of $O_n$ as the coefficients in the series
\begin{equation}
    \om_0^+(t)+\om_0^-(t)= \sum_{n=0}^\infty O_{2n-1}t^{2n-1}.
\end{equation}
Using resummation formulae, facts 1 and 2,  \eqref{eq: Basic asymptotic, no log} and \eqref{eq: asymptotic log}, one then finds:
\begin{equation}
\begin{split}
    \CR_{\lambda,s}^{(0)} &=  \sum_{n\in \bb{Z}-s>0}G_\lambda(n\Theta) =\sum_{m=0}^{\infty}G_\lambda((m+s)\Theta)\\
    &\sim \frac{1}{\Theta}\int_0^{\infty}G_{\lambda}(x)dx - \pi i\sum_{n=1}^\infty \zeta(1-2n,1-s)O_{2n-1}\Theta^{2n-1} -\delta_{\lambda,0}\delta_{s,0}\log\left(\frac{i\Theta}{\a_1\t}\right) \\
    &+\log(2)(\delta_{\lambda,0}\delta_{s,\frac{1}{2}}-\delta_{s,0}\delta_{\lambda,\frac{1}{2}}.)
\end{split}
\end{equation}
Let us evaluate the integral
\begin{equation}
\begin{split}
    \int_0^{\infty}G_{\lambda}(x)dx &= \int_0^{\infty}\log(1-e^{2\pi i\lambda }e^{-2\pi i \om_0^{-}(x)}) +\log(1-e^{2\pi i\lambda }e^{-2\pi i \om_0^{+}(x)})dx\\
    &=2\pi i\int_0^{\om_0^-(\infty)}\frac{x(\om_0^{-})}{e^{2\pi i \lambda}e^{2\pi i \om_0^{-}}-1}d\om_0^{-}+2\pi i\int_0^{\om_0^+(\infty)}\frac{x(\om_0^{+})}{e^{2\pi i \lambda}e^{2\pi i \om_0^{+}}-1}d\om_0^{+}.
\end{split}
\end{equation}
Recall that the roots $\om_0^{\pm}(x)$ satisfy
\begin{equation}
    \pm\sum_{j=1}^{N}\t^j\a_j(\pm1)^{j}(\om^{\pm}_0(x))^j=x,
\end{equation}
which we can use to define $x(\om_0^{-})$ and $x(\om_0^{+})$ in each integral. We can also perform the substitution $-iw = \om_0^{-}$ and $-iw = \om_0^+$ in each integral. We obtain
\begin{equation}
    -2\pi \sum_{j=1}^{N}\t^j\a_ji^j\left[(-1)^{j+1}\int_0^{i\om_0^{-}(\infty)}\frac{w^j}{e^{2\pi i \lambda}e^{2\pi w}-1}dw+\int_0^{i\om_0^{+}(\infty)}\frac{w^j}{e^{2\pi i \lambda}e^{2\pi w}-1}dw\right].
\end{equation}
Both are integrals on the right hand side of the complex $w$ plane, so never cross the imaginary axis where the poles of the integrand are, and so we can deform contours on which the integrals are defined to become the real line. It is clear then that the terms with even $j$ will cancel, and the terms with odd $j$ add up. We find
\begin{equation}
    4\pi \sum_{n=1}^K \t^{2n-1}\a_{2n-1}i^{2n-1}\int_0^{\infty}\frac{w^{2n-1}}{e^{2\pi i \lambda}e^{2\pi w}-1}dw = -2\pi i \sum_{n=1}^{K}\t^{2n-1}\a_{2n-1}\zeta(1-2n,1-\lambda),
\end{equation}
where we've used the integral definition of the Hurwitz zeta function. In the above, $2K-1$ is the spin of the highest odd-spin charge inserted in the GGE. Bringing everything together we find that the sum over the roots is given asymptotically by
\begin{equation}\label{eq: Expansion of SCR final}
\begin{split}
        \CR_{\lambda,s}&\sim -\theta2\pi i \sum_{n=1}^{K}\t^{2n-1}\a_{2n-1}\zeta(1-2n,1-\lambda) - \pi i\sum_{n=1}^\infty \zeta(1-2n,1-s)O_{2n-1}\theta^{1-2n}\\
        &-\delta_{\lambda,0}\delta_{s,0}\log\left(\frac{i}{\a_1\t\theta}\right) +\log(2)(\delta_{\lambda,0}\delta_{s,\frac{1}{2}}-\delta_{s,0}\delta_{\lambda,\frac{1}{2}})
\end{split}
\end{equation}
where we've remembered that $\theta = \Theta^{-1}$.
\subsection*{Piecing together}
Recall that we are analysing \eqref{eq: The equation we're expanding}, which we restate for clarity:
\begin{equation}
    \CB_{\lambda,s} +\log(\theta)\delta_{\lambda,0}\delta_{s,0} = \CI_{\lambda,s} +\CR_{\lambda,s} + \CC_{\lambda,s}\theta + \CD_{\lambda,s}.
\end{equation}
Using all of the previous results for asymptotic expansions around $\theta\to \infty$, \eqref{eq: Expansion of SCB Final}, \eqref{eq: Expansion of SCI Final} and \eqref{eq: Expansion of SCR final}, we find that the integration constants can only be:
\begin{equation}
\begin{split}
\CC_{\lambda,s}&=2\pi i \sum_{n=1}^K \t^{2n-1}\a_{2n-1}\zeta(1-2n,1-\lambda)\\
\CD_{\lambda,s}&=\delta_{\lambda,0}\delta_{s,0}\log\left(\frac{i}{\a_1\t}\right)+\log(2)(\delta_{s,0}\delta_{\lambda,\frac{1}{2}}-\delta_{s,\frac{1}{2}}\delta_{\lambda,0})
\end{split}
\end{equation}
Now recall that our original GGE was defined at $\theta = 1$, and that we have the relation \eqref{eq: Definition of SCB}
\begin{equation}
    \CB_{\lambda,s} = \log(\Psi(\theta|\mathbf{\a}))+2\pi i\sum_{n=1}^{K}\t^{2n-1}\a_{2n-1}\zeta(1-2n,1-\lambda).
\end{equation}
 We can also set $\a_1=1$ for simplicity, and obtain a final answer:
\begin{equation}
\begin{split}
\Psi_{\lambda,s}(\t,\a)=\frac{2^{\delta_{s,0}\delta_{\lambda,1/2}}}{2^{\delta_{s,1/2}\delta_{\lambda,0} }}\left(\frac{i}{\t}\right)^{\delta_{\lambda,0}\delta_{s,0}}\prod_{\pm} & \exp \biggl \{  \int_{0}^\infty (\log(1-e^{2\pi i s}e^{2\pi i ( \sum_{n=1}^N(\pm1)^{n+1}\a_n\t^{n}x^n)})) dx\biggr \}\\
&\times\left(\prod_{n \in \bb{Z}-s}\prod_{ \om^{\pm}_l(n) \in {\h}^{\raisebox{3pt}{$\scriptscriptstyle +$}}}(1-e^{2\pi i \lambda}e^{2\pi i  \om_l ^ \pm(n)})\right),
\end{split}
\label{eq:final1}
\end{equation}
where the roots $\om_l^{\pm}(\kappa)$ satisfy the polynomial equation (with $\alpha_1=1$)
\begin{equation}
    \om = \om_l^{\pm}(\kappa): \quad\pm\sum_{j=1}^N(\pm1)^{j}(\t^j\a_j\om^j)=\kappa
    \label{eq:final2}
\end{equation}
To arrive at this answer, we had to work in the restricted domain \eqref{eq: Restricted Domain}, but we can see that it is still well defined and convergent in the domain \eqref{eq: Expanded domain}, and so by analytic continuation, this must also be the transformation in that expanded domain too.
What we have is an answer starting with $\tau$ on the left hand side, and ending with $\that = -1/\t$ on the right hand side. We could have done the whole proof starting with $\that$, which would then give us the answer:
\begin{equation}\label{eq: Final Transformation}
\boxed{\begin{split}
\Psi_{\lambda,s}(\that,\a)=\frac{2^{\delta_{s,0}\delta_{\lambda,1/2}}}{2^{\delta_{s,1/2}\delta_{\lambda,0} }}\left(\frac{\t}{i}\right)^{\delta_{\lambda,0}\delta_{s,0}}\prod_{\pm} & \exp \biggl \{  \int_{0}^\infty (\log(1-e^{2\pi i s}e^{2\pi i ( \sum_{n=1}^N(\pm1)^{n+1}\a_n\that^{n}x^n)})) dx\biggr \}\\
&\times\left(\prod_{n \in \bb{Z}-s}\prod_{ \om^{\pm}_l(n) \in {\h}^{\raisebox{3pt}{$\scriptscriptstyle +$}}
}(1-e^{2\pi i \lambda}e^{2\pi i  \om_l ^ \pm(n)})\right),
\end{split}}
\end{equation}
with roots satisfying  (with $\alpha_1=1$)
\begin{equation}\boxed{
\om= \om_l^{\pm}(\kappa): \quad\pm\sum_{j=1}^N(\pm1)^{j}(\that^j\a_j\om^j)=\kappa\;.
\label{eq:roots}}
\end{equation}
We have chosen to do this for two reasons. The first is that in the next section, we will perform perturbative checks by evaluating thermal correlation functions at $\that$ and getting transformed expressions in terms of just $\t$. The second is purely stylistic, we find it more intuitive to evaluate a function at the $S$ transformed point $\that$ and find out what happens in terms of $\t$ rather than the other way around. For a sanity check, one can set all of the chemical potentials $\a_n=0$ for $n > 1$. In that case, one would recover precisely the normal character transformation formulae \eqref{eq: T trasform}, \eqref{eq: S trasform}, \eqref{eq: T pseudo character}, \eqref{eq: S pseudo character}.

\section{Asymptotic Analysis of Symplectic Fermion GGE}\label{sec: Asymptotic Analysis of Symplectic Fermion GGE}
In this section, we will perform an asymptotic analysis of the modular transformation of the GGE obtained in the previous section, in the regime where the chemical potentials go to zero. This enables a comparison with the result in \cite{Downing:2025huv}. As a sanity check, we also compare this result against perturbative analysis. We also discuss the connection with the $c=\frac{1}{2}$ free fermion result from \cite{Downing:2021mfw}.

For simplicity, throughout the whole of this section, we will assume that $\that$ is purely imaginary.

\subsection{Asymptotic Check with $\Q_{2}$}
Let us begin this section by reiterating the main result for a GGE with just the spin-2 charge inserted. With $\that = -1/\t$,
\begin{equation}
\begin{split}
    \Tr_{\lambda,s}(\hat{q}^{L_0-c/24}e^{\mu_2\Q_2})= \frac{2^{\delta_{s,0}\delta_{\lambda,1/2}}}{2^{\delta_{s,1/2}\delta_{\lambda,0} }}\left(\frac{\t}{i}\right)^{\delta_{\lambda,0}\delta_{s,0}}\prod_{\pm}&\exp \biggl \{  \int_{0}^\infty (\log(1-e^{2\pi i s}e^{2\pi i (\hat{\t} x \pm \hat{\t}^2x^2\a_2)})) dx\biggr \}\\
        &\times \left(\prod_{n \in \bb{Z}-s}\prod_{ \om^{\pm}_l(n) \in \bb{H}^+}(1-e^{2\pi i \lambda}e^{2\pi i  \om_l ^ \pm(n)})\right),
\end{split}
\end{equation}
where we have that the roots $\om_l^{\pm}(\kappa)$ satisfy:
\begin{equation}
    \om= \om_l^{\pm}(\kappa): \qquad (\hat{\t}\om) \pm \a_2 (\hat{\t} \om)^2 = \kappa,
\end{equation}
and we also have that $\mu_2 =: 2\pi i \that^2 \a_2$. First we will perform an asymptotic expansion of this result around $\mu_2 \to 0$. Then we will check that this is what we expect from perturbative analysis.

The problem of asymptotically expanding the above result is broken down into two parts. The first part is dealing with the integral expression to obtain ground state energies. The second part is dealing with the roots of the polynomials appearing in the infinite product, corresponding to excited state energies. We do not need to deal with the constant prefactor since this is just a compact encoding of the $S$-transformation of the symplectic fermion characters \eqref{eq: T trasform}, \eqref{eq: S trasform}, \eqref{eq: T pseudo character}, \eqref{eq: S pseudo character}.

\subsubsection{Ground State Energies} The integral at hand is
\begin{equation}
    \prod_\pm\exp \biggl \{  \int_{0}^\infty (\log(1-e^{2\pi i s}e^{2\pi i (\hat{\t} x \pm \hat{\t}^2x^2\a_2)})) dx\biggr \} =: q^{h_0^{(s)}(\t,\mu_2)},
\end{equation}
where $q = e^{2\pi i \t}$. Perform now the substitution $u = -2\pi i \that x$. Then one finds that
\begin{equation}
    h_0^{(s)}(\t,\mu_2) = - \frac{1}{4\pi ^2}\sum_{\pm}\int_0^{\infty} \log(1 - e^{2\pi is}e^{-u \pm \mu_2(\frac{u \t}{2\pi i})^2})du.
\end{equation}
Now on each integral we perform another substitution. That is
\begin{equation}
    u=: tf_{\pm}(t), \quad -(t f_{\pm}(t)) \pm \mu_2 \left(\frac{\t}{2\pi i}\right)^2 (t f_{\pm}(t))^2 = - t.
\end{equation}
We choose $f_{\pm}(t)$ such that
\begin{equation}
\begin{split}
t f_{\pm}(t)\log(1-e^{2\pi is}e^{-t})&\to 0\quad\text{as} \quad t \to 0,\\
t f_{\pm}(t)\log(1-e^{2\pi is}e^{-t})&\to 0\quad\text{as} \quad t \to \infty,\\
t f_{\pm}(t)\to \infty \quad&\text{as} \quad t \to \infty.\\
\end{split}
\end{equation}
Such an $f_{\pm}(t)$ can be found explicitly
\begin{equation}
    f_{\pm}(t) = \mp \frac{2 \pi ^2}{\mu_2 \t^2 t}\left(1- \sqrt{1\pm \frac{t\mu_2\t^2}{\pi^2}}\right).
\end{equation}
Then the integral becomes
\begin{equation}
    h_0^{(s)}(\t,\mu_2) = \frac{e^{2\pi is}}{4\pi ^2}\int_0^{\infty} \frac{t}{e^t - e^{2\pi is}}(f_+(t) + f_-(t))dt.
\end{equation}
Now we perform a series expansion of the integrand and integrate each term. Since $\mu_2$ is constrained to be purely imaginary, we find that such an expansion is at best an asymptotic expansion. The result is
\begin{equation}\label{eq:h0(s)(t,mu2)}
    h_0^{(s)}(\t,\mu_2)\sim -\frac{1}{\mu_2 \tau^2}\sum_{m=0}^{\infty} \frac{2^{-4m-2}(2\pi i)^{2m+2}}{(2m+1)} \binom{4m}{2m}\left(\frac{\mu_2\t^2}{\pi^2}\right)^{2m+1}\mathbf{c}_{2m+1}(s),
\end{equation}
where ${\bf c}_n(\lambda)$ is defined in \eqref{eq: THE CHARGES}. We can now move on to the excited state energies.

\subsubsection{Excited State Energies}

The excited state energies appear in the product
\begin{equation}
    \prod_{n \in \bb{Z}-s}\prod_{ \om^{\pm}_l(n) \in \bb{H}^+}(1-e^{2\pi i \lambda}e^{2\pi i  \om_l ^ \pm(n)}).
\end{equation}
The roots satisfy
\begin{equation}
    \om = \om_l^{\pm}(\kappa): \qquad (\hat{\t} z) \pm \a_2 (\hat{\t} \om)^2 = \kappa.
\end{equation}
What is convenient for us is that this is just a quadratic equation and we can solve this directly and simply. As stated earlier, for simplicity, we will assume $\that$ is pure imaginary $\that = i \that_2$ (so that $\t = \frac{i}{\that_2}$). The $(+)$ equation has two solutions:
\begin{equation}
    \om^{+}_{0} =-\frac{i \left(\sqrt{4 \alpha _2 \kappa +1}-1\right)}{2 \alpha _2 \that _2}, \quad \om^{+}_{1} = \frac{i \left(\sqrt{4 \alpha _2 \kappa +1}+1\right)}{2 \alpha _2 \that _2} 
\end{equation}
and the $(-)$ equation also has two solutions:
\begin{equation}
   \om^{-}_{0} = \frac{\sqrt{4 \alpha _2 \kappa -1}-i}{2 \alpha _2 \that _2},\quad \om^{-}_{1} =-\frac{\sqrt{4 \alpha _2 \kappa -1}+i}{2 \alpha _2 \that _2}.
\end{equation}
We choose these solutions so that $\om_0$ is the solution continuously connected to zero when $\kappa=0$. Performing a series expansion of each in $\a_2$, we find that only $\om_0^\pm$ does not diverge as $\a_2 \to 0$, since 
\begin{equation}
    \om_1^{\pm}(\kappa) \sim \frac{\mp i }{\that_2\a_2} + O(\a_2) \quad \text{as}\quad \a_2 \to 0. 
\end{equation}
It is clear then that $\om_1$ is either not in the upper half plane, or it is in the upper half plane but diverges as $\a_2\to 0$ so that its contribution to the product will be
\begin{equation}
    (1-e^{2\pi i \lambda}e^{2\pi i \om_1}) \to 1,
\end{equation}
so there will be no overall contribution from $\om_1^\pm$ in the $\a_2\to 0$ limit.

It is not too difficult then to calculate
\begin{equation}
    2\pi i\om_0^{\pm}(n) \sim 2 \pi i \t n + n f(\mp n),
\end{equation}
where
\begin{equation}\label{eq: f(omega)}
    f(\kappa) = \sum_{m=0}^{\infty}\frac{i 2^{m+2} \pi ^{-m} \tau  \kappa^{m+1} \left(\frac{1}{2}\right)_{m+1} \left(-i \mu_2  \tau ^2\right)^{m+1}}{(2)_{m+1}}.
\end{equation}
The product asks for solutions in the upper-half plane which occurs if $\kappa>0$, so we find that the product becomes
\begin{equation}
      \prod_{n\in \bb{N}-s}^{\infty}(1+q^ne^{n f(n)})(1+q^{n}e^{nf(-n)}).
\end{equation}
\subsubsection{The Complete Asymptotic Expression}
We can carefully massage the expressions \eqref{eq:h0(s)(t,mu2)} and \eqref{eq: f(omega)} and see that they are exactly expressions that we would expect from a GGE with infinitely many higher spin bilinear charges inserted. Indeed, we can see that our transformation in the asymptotic limit is the following GGE:
\begin{equation}
    \Tr_{\lambda,s}(\hat{q}^{L_0-c/24}e^{\mu_2\Q_2}) \sim \frac{2^{\delta_{s,0}\delta_{\lambda,1/2}}}{2^{\delta_{s,1/2}\delta_{\lambda,0} }}\left(\frac{\t}{i}\right)^{\delta_{\lambda,0}\delta_{s,0}}\Tr_{s,\lambda} (q^{L_0-c/24}e^{\mu_2 \t^3 \text{{\boldmath{$\SCQ$}}}_2}),
\end{equation}
where $\text{{\boldmath{$\SCQ$}}}_2$ is defined by:
\begin{equation}\label{eq: SCQ 2}
\text{{\boldmath{$\SCQ$}}}_2 
        = \sum_{k=0}^{\infty}\left(\left(\frac{\mu_2\t^2}{4\pi i}\right)^{k}\frac{1}{k!}\frac{2^{k+1}(2k+1)!}{(k+2)!}\Q_{k+2}\right).
\end{equation}
\subsubsection{Perturbative Check}
We can perform a perturbative check by expanding directly in terms of $\mu_2$ the expression:
\begin{equation}
    \Tr_{\lambda,s}(\hat{q}^{L_0-c/24}e^{\mu_2\Q_2}),
\end{equation}
and transforming each term separately, and then picking out the terms lowest order in $\t$ at each order in $\mu_2$. As emphasised before, this expansion is asymptotic since $\mu_2$ is confined to be purely imaginary. We will for simplicity in this section suppress the subscripts $\lambda,s$ and all modular $S$-matrices and leave them implicit. The expansion is (recalling the definition of the expectation value in \eqref{eq: Def Exp Value +})
\begin{equation}
    \Tr(\hat{q}^{L_0-c/24}e^{\mu_2\Q_2}) \sim \sum_{k=0}^{\infty} \frac{\mu_2^k}{k!}\vev{\Q_2^k}(\that).
\end{equation}

One can find the transformations of $\vev{\Q_2^k}$ in various ways. In \cite{Downing:2025huv}, several methods were used.
One is the use of Zhu's recursion relation \cite{Zhu1996}; another is to compute $\vev{\Q_2^k}$ via a modular linear differential operator by looking at the $q$-series expansion of  
\begin{equation}
    \vev{\Q_2^k} = \frac{\del^k}{\del \mu_2^k}\biggl|_{\mu_2=0}\vev{e^{\mu_2 \Q_2}},
\end{equation}
and then matching perturbatively against the $q$-series expansion of an ansatz for a quasi-modular differential operator of appropriate weight and depth, analogous to \cite{Maloney:2018hdg}. Once you have such an expression, you can compute the $S$-transformation relatively easily, and figure out what thermal correlation functions of what charges correspond to the resulting expression. Finally, we explored yet another method involving Toda field theory. Whatever method one employs, the result is the same.

By the commutation relation \eqref{eq: Even Commuter},
by interchanging the $\pm$ labels on the basis states, 
the contributions to the trace
either cancel in pairs or vanish identically, and so
\begin{equation}
    \vev{\Q_2^{2n-1}}=0 \quad n \in \bb{N}.
\end{equation}
A general expression for the even powers is not available to us, but we can go case by case. We find for the first two cases that\footnote{We're suppressing the modular $S$-matrix appearing in these terms to remove clutter and make clearer the charges appearing in the transformation}
\begin{equation}
    \vev{\Q_2^2}(\that) = - \frac{2 i \t^5}{\pi}\vev{\Q_3}(\t) + \t^6\vev{\Q_2^2}(\t),
\end{equation}
\begin{equation}\label{eq: Q24}
    \vev{\Q_2^4}(\that) =  \frac{42i\t^9}{\pi ^3}\vev{\Q_5}(\t)-\frac{30 \t^{10}}{\pi^2}\vev{\Q_2\Q_4}(\t)-\frac{12 \t^{10}}{\pi^2}\vev{\Q_3^2}(\t) + O(\t^{10}).
\end{equation}
It is clear from these expressions that there will appear even spin charges in the transformed GGE once we resum it into an exponential. However, since the one-point functions of even spin charges vanish, it is not immediately clear whether the charges will appear with a positive or negative sign. There are two points to highlight in connection with this.

The first is that the Symplectic Fermion GGE, in the untwisted and half-twisted sectors, enjoys, for each even spin bilinear charge inserted in the GGE, a $\bb{Z}_2$ symmetry, $\mu_{2n}\to -\mu_{2n}$,
\begin{equation}
    \Tr(q^{L_0-c/24} e^{...+\mu_{2n}\Q_{2n}+...})
    =
    \Tr(q^{L_0-c/24} e^{...-\mu_{2n}\Q_{2n}+...}).
\end{equation}
This fact can be confirmed by inspection of the GGE \eqref{eq: GGE general}. So either way, the GGE will be the same function.

The second point is that if we were instead considering this GGE in a generic theory, the one-point functions would not vanish. In that case, given our choice of representative for the modular $S$ transformation \eqref{eq: Representative for S}, one finds that the transformation of any charge with even spin $2n$, denoted $I$, the zero-mode of an odd-weight quasi-primary field \eqref{eq: I = hat J 0}, is
\begin{equation}
    \vev{I}\!\left(- \tfrac{1}{\t}\right)
    =
    \t^{2n-1} \vev{I} .
\end{equation}
Indeed, so as to make the connection consistent with the conjecture in our companion work \cite{Downing:2025huv}, we should take the positive sign for the transformation of $\Q_2$, and based on \eqref{eq: Q24}, the relative sign is fixed for how $\Q_4$ appears, and so on. We expect that the first terms appearing in the exponential in the GGE to be 
\begin{equation}
    \Tr(\hat{q}^{L_0-c/24}e^{\mu_2\Q_2}) \sim  \Tr(q^{L_0-c/24}e^{\mu_2\t^3 \Q_2-\frac{\mu_2^2 i \t^5}{\pi}\Q_3 -\frac{5 \mu_2 ^3 \tau ^7}{4 \pi ^2}\Q_4+\frac{7 i \mu_2^4   \tau ^9}{4 \pi ^3}\Q_5+...}).
\end{equation}

These  charges are precisely what one would find in the first terms of $\text{{\boldmath{$\SCQ$}}}_2$ in \eqref{eq: SCQ 2}.

It is worth mentioning that in our companion paper \cite{Downing:2025huv}, we propose a conjecture for the general all-order in $\mu_2$ expression for the asymptotic transformation of the GGE with the spin-2 charge inserted. In that paper, the conjecture is checked to be true without specifying a model up to $O(\mu_2^7)$. We show in that work that the asymptotic result \eqref{eq: SCQ 2} matches that conjecture to all orders in $\mu_2$ at $c=-2$.
\subsection{Asymptotic Check with $\Q_{2n-1}$}\label{subsubsec:Asymptotic Check with Q 2n-1}
 The Symplectic Fermion GGE with KdV charges is simply the square of the Free Fermion GGE with KdV charges. In this way, our exact transformation with odd-spin charges matches exactly the result proved in \cite{Downing:2023lop}. The free fermion has two sectors, the half-integer moded Neveu-Schwarz sector, and the integer moded Ramond sector. If $\CO$ is an operator in the free fermion, then the following traces are defined
\begin{equation}
\begin{split}
    \Tr^{\text{FF}}_{0,1/2}(\CO) = \Tr_{\text{R},+}(\CO),\quad&\Tr^{\text{FF}}_{1/2,1/2}(\CO) = \Tr_{\text{NS},+}(\CO),\quad\Tr^{\text{FF}}_{1/2,0}(\CO) = \Tr_{\text{NS},-}(\CO)\\
    & \Tr_{0,0}^{\text{FF}}(\CO) = \Tr_{\text{R},-}(\CO)
\end{split}
\end{equation}
where the trace $\Tr_{\text{R},-}$ is taken over the Ramond sector but omitting the zero mode $\psi_0$. Denote by $\I_{2n-1}^{\text{FF}}$ the KdV charges in the theory of the Free Fermion \cite{Downing:2021mfw}. Then one can easily see when we insert the first $N$ KdV charges into the GGE we have the relations
\begin{equation}\label{eq: Squaring Charges}
    \Tr_{\lambda,s}(e^{\sum_{k=1}^N\mu_{2k-1}\Q_{2k-1}}) =  \left[\Tr^{\text{FF}}_{\lambda,s}(e^{\sum_{k=1}^N\mu_{2k-1}\I^{\text{FF}}_{2k-1}})\right]^2.
\end{equation}
 In \cite{Downing:2021mfw}, they calculated what the asymptotic expression for the transformed GGE should be for the charge $\I^{\text{FF}}_{2m-1}$. This transformation was done for a generic $\slz$ transformation:
\begin{equation}
    \that = \frac{a \t  + b}{ c \t +d}.
\end{equation}
Their result was as follows:
\begin{equation}\label{eq: Asymptotic Downing Result}
    \Tr^{\text{FF}}(\hat{q}^{L_0-\frac{c}{24}}e^{\mu \I^{\text{FF}}_{2m-1}}) \sim \Tr^{\text{FF}} (e^{\sum_{p=0}^{\infty} \mu_{2p+1}^{(m)}\I^{\text{FF}}_{2p+1}}).
\end{equation}
Here we have suppressed indices and resulting prefactors associated to the modular transformation to avoid clutter. The terms $\mu_{2p+1}^{(m)}$ are defined below, noting that if $p$ is divisible by $m-1$ then $n := p/(m-1)$:
\begin{equation}
\begin{split}
    \mu_{1}^{(m)}&=2\pi i \t\\
    \mu_{2p+1}^{(m)}&= \delta_{p,n(m-1)} \,\, \frac{\mu^n}{n!(2\pi i)^{n-1}}\frac{(n(2m-1))!c^{{n-1} }}{(2n(m-1)+1)!}(c \t + d)^{n(2m-1)+1}.
\end{split}
\end{equation}
That is, the terms vanish when $m-1$ does not divide $p$.
One concludes that the result for the symplectic fermion is the same. 
\begin{equation}\label{eq: Asymptotic SImp}
\Tr(\hat{q}^{L_0-\frac{c}{24}}e^{\mu \Q_{2m-1}}) \sim \Tr(e^{\sum_{p=0}^{\infty} \mu_{2p+1}^{(m)}\Q_{2p+1}}).
\end{equation}
The significance of this result is that it resolves a conjecture posed in \cite{Downing:2024nfb}. There, it was conjectured that the only two values of the central charge where multipoint thermal correlation functions of KdV charges  close under modular transformations are 
$c=1/2$ and $c=-2$. This was because at generic central charge the two-point KdV thermal correlation function S-transforms as
\begin{equation}\label{eq: c+2 term}
\vev{\left(\I^{\text{KdV}}_{3}\right)^2}(\tfrac{-1}{\t})= \t^8\vev{\left(\I^{\text{KdV}}_{3}\right)^2}(\t) -\frac{i \t^7}{\pi} \left(4 \vev{\I_5^{\text{KdV}}}+ \frac{5(c+2)}{54}\vev{\J_5}\right).
\end{equation}
$\J_5$ is the zero-mode of a weight-6 quasi-primary field which is linearly independent of the current that gives the KdV charge. At $c=1/2$ this quasi-primary field differs from the KdV current by a null field, and at $c=-2$ its contribution 
vanishes directly.

From \eqref{eq: Asymptotic Downing Result} and  \eqref{eq: Asymptotic SImp} we see that, at these central charge values, multi-point thermal correlators of the KdV charges always map to multi-point thermal correlation functions of KdV charges. From \eqref{eq: c+2 term} we see that this can only happen at $c=1/2$ and $c=-2$.

\section{Relation to $W_n$ algebras and other hierarchies}
\label{sec:Wn}
\label{subsec: Realisation of Hierarchies}

As we have already said, the symplectic fermion's relation to the KdV and Boussinesq hierarchy were shown a long time ago in \cite{DiFrancesco:1991xm}. 
In this section we demonstrate that the bilinear quasi-primary fields introduced earlier realise both the KdV hierarchy and the Boussinesq hierarchy and make some comments on the relation to the hierarchies of other $W$ algebras.

\subsection{The KdV Hierarchy} 
\label{subsec:KdV}

The KdV charges are defined as zero-modes of even-spin quasi-primary fields constructed solely from the stress tensor:
\begin{equation}
    \I_{2n-1}^{\text{KdV}} = \int_{0}^{2\pi i}\!\frac{dw}{2\pi i}\, J^{\text{KdV}}_{2n}(w),
\end{equation}
where the charges $\I_{2n-1}^{\text{KdV}}$ mutually commute. These are defined at generic central charge \cite{Bazhanov:1994ft}, and below we have written the first two currents in the series, having added ourselves the total derivative pieces necessary to make these fields quasi-primary.
\begin{equation}
\begin{split}
    J^{\text{KdV}}_2 &= T(z),\\
    J^{\text{KdV}}_4 &= \Lambda(z) = (TT)(z)-\tfrac{3}{10}\,\partial^2 T(z).
\end{split}
\end{equation}

In \cite{DiFrancesco:1991xm} it was already shown that the KdV hierarchy can be expressed in terms of bilinears of symplectic fermion fields. What they call $\phi$ and $\psi$, we have denoted by $\chi^-$ and $\chi^+$. The result is that, at $c=-2$, they find a set of weight $2n$ fields in the Virasoro algebra, whose spin $2n-1$ zero-modes commute with one another, expressed simply as left normal orderings of the stress tensor $(\overleftarrow{T^n})$.  
\begin{equation}
    (\overleftarrow{T^n}) = (\cdots(((TT)T)T)\cdots T)\quad(n\ \text{factors}).
\end{equation}
Since the KdV hierarchy is the unique set of such fields/charges, they conclude that these left normal ordered stress tensors correspond to the KdV currents. That is, up to total derivatives,
\begin{equation}
    c=-2:\quad(\overleftarrow{T^n}) \propto J^{\text{KdV}}_{2n}
\end{equation}
In terms of the symplectic fermion, they show that these KdV currents can be written simply as bilinear fields. 
\begin{equation}
    (\overleftarrow{T^n}) = \frac{n}{2^n} \left((\del^{2n-2}\chi^- \chi^+) + (\chi^- \del^{2n-2}\chi^+) \right).
\end{equation}

Based on \eqref{eq: Bilinear field, but Ank}, we can write 
\begin{equation}
    B_{2n} = \frac{A_{2n,0}2^n}{n} (\overleftarrow{T^n}) + \sum_{k=1}^{2n-3}A_{2n,k}(\del^{2n-2-k}\chi^-\del^k\chi^+).
\end{equation}
It is easy to check from this expression that $B_{2n}$ is proportional to $(\overleftarrow{T^n})$ up to total derivatives.

\subsection{The $W_3$ algebra and the  Boussinesq Hierarchy}
\label{subsec:Boussinesq}

The Boussinesq hierarchy is associated with the $W_3$ algebra.
The chiral algebra of the $\CW(1,2)$ triplet model consists of all bosonic fields constructed from the symplectic fermions, and in particular includes (as a subalgebra) the $W_3$ algebra, also known as the ``Singlet Algebra" at $c=-2$. More precisely, this is a $U(1)$ orbifold of the Triplet algebra.

The generators are\footnote{Note that this is neither of the two standard normalisations of the field $W(z)$, the original normalisation of 
\cite{Zamolodchikov:1985wn} nor that of \cite{Cardy:1988tk}, where the coefficient of the sixth order pole in $W(z)W(w)$ in \eqref{eq: W3 algebra} would be $-2/3$ and $-1/6$ respectively, but is chosen to make the commutators with the symplectic fermion modes \eqref{eq:modecomms} simple.
}
\begin{equation}\label{eq: W_3 symmetry}
    T(z) = (\chi^-\chi^+)(z), \quad 
     W(z) = \Big((\del \chi^-\chi^+) - \tfrac{1}{2}\del(\chi^-\chi^+)\Big)(z).
\end{equation}
Their operator product expansions (OPEs) realise the $W_3$ algebra:
\begin{equation}\label{eq: W3 algebra}
\begin{split}
T(z)T(w) &\sim \frac{-1}{(z-w)^4}+\frac{2T(w)}{(z-w)^2}+\frac{\del T(w)}{z-w},\\
T(z) W(w) &\sim \frac{3 W(w)}{(z-w)^2}+\frac{\del W(w)}{z-w},\\
 W(z) W(w) &\sim \frac{-1}{(z-w)^6}+\frac{3T(w)}{(z-w)^4}+\frac{3}{2}\frac{\del T(w)}{(z-w)^3}\\
&\quad + \frac{1}{(z-w)^2}\!\left(4(TT)(w)-\tfrac{3}{4}\del^2T(w)\right) 
 + \frac{1}{z-w}\!\left(2\del(TT)(w)-\tfrac{1}{2}\del^3T(w)\right).
\end{split}
\end{equation}
The OPEs with the Symplectic Fermions simplify:
\begin{equation}
    T(z)\chi^{\pm}(w) \sim \frac{\chi^{\pm}(w)}{(z-w)^2}+\frac{\del\chi^{\pm}(w)}{z-w},
\end{equation}
\begin{equation}
    W(z)\chi^{\pm}(w) \sim \mp\frac{\chi^{\pm}(w)}{(z-w)^3}\mp\frac{3}{2}\frac{\del\chi^{\pm}(w)}{(z-w)^2}\mp\frac{\del^2\chi^{\pm}(w)}{z-w}.
\end{equation}
Equivalently, in terms of modes:
\begin{equation}
\begin{split}
    [L_m,\chi^{\pm}_n] &= -n\,\chi^{\pm}_{m+n},\\
    [W_m,\chi^{\pm}_n] &= \mp \tfrac{1}{2}n(m+2n)\chi^{\pm}_{m+n}.
\end{split}
\label{eq:modecomms}
\end{equation}

We will not discuss the details of the representation theory of the singlet algebra here, although we direct the reader to both our earlier brief discussion of the generic twisted sectors of the Symplectic Fermion to which they are related \ref{subsec: Expressions for the charges}, and to \cite{Creutzig:2013hma} for details. We will also briefly discuss $W_3$ representation theory in Appendix \ref{app: w0 lambda 0}. At the moment, we just note that the generic twisted representations do not enter the modular invariant partition function of the triplet model \eqref{eq: triplet Modular Invariant.} considered in this work. 

In analogy with the KdV case, the Boussinesq charges are associated with spins that are integers not divisible by $3$, but in the $W_3$ algebra rather than Virasoro,
\begin{equation}
    \I_{n}^{\text{Bou.}} = \int_{0}^{2\pi i}\!\frac{dw}{2\pi i}\, J_{n+1}(w),
    \qquad n = 1,2 \pmod 3.
\end{equation}
The first three Boussinesq currents are given at generic central charge by \cite{Bazhanov:2001xm,Kupershmidt:1989bf}
\begin{equation}
\begin{split}
    J^{\text{Bou.}}_2(z) &= T(z),\\
    J^{\text{Bou.}}_3(z) &= W(z),\\
    J^{\text{Bou.}}_5(z) &= (TW)(z) - \tfrac{3}{14}\,\partial^2 W(z).
\end{split}
\end{equation}
where again we have added total derivative terms to make the fields quasi-primary without affecting the integrals of motion. We have already shown that the first two are bilinear fields, \eqref{eq: W_3 symmetry}:
\begin{equation}
    T(z) = B_2(z), \qquad W(z) = B_3(z).
\end{equation}
For the next charge, applying the state-operator correspondence as before yields
\begin{equation}
\begin{split}
    J^{\text{Bou.}}_5 &= \tfrac{5}{84}(\partial^3\chi^-\, \chi^+) 
    - \tfrac{5}{14}(\partial^2\chi^-\, \partial \chi^+) 
    + \tfrac{5}{14}(\partial \chi^-\, \partial^2\chi^+) 
    - \tfrac{5}{84}(\chi^-\, \partial^3\chi^+) \\
    &= \tfrac{5}{6}\,B_5.
\end{split}
\end{equation}
Hence the first three Boussinesq currents are proportional to the bilinear fields, and therefore so are their charges. Explicit checks up to spin $13$ give
\begin{equation}
    \I_{n}^{\text{Bou.}} = \Q_{n}, 
    \qquad n = 1,2 \pmod 3,\quad n \leq 13.
\end{equation}

We therefore expect that the full Boussinesq hierarchy is realised at $c=-2$, with currents constructed from bilinear Symplectic Fermion fields. It should be possible to mirror the analysis of \cite{DiFrancesco:1991xm} and show that there exists a set of fields composed of $T(z)$ and $W(z)$ such that the spins of their charges are exactly $1,2 \pmod3$ and when written in terms of symplectic fermions they are proportional to the bilinear fields presented here.

\subsection{The relation of KdV and Boussinesq charges}

The KdV and Boussinesq charges are distinct for general values of $c$, and in particular the Boussinesq hierarchy does include $\Lambda_0$, 
a charge of weight 3, whereas the KdV hierarchy does; and conversely the Boussinesq hierarchy has a weight two charge $W_0$ which the KdV hierarchy does not. If one computes their commutator, one gets
\begin{align}
[W_0,\Lambda_0]
= -\frac{1}5(
12 (T' W) - 8  (W'T) + W''')_0
\;,
\label{eq:W0L0}
\end{align}
and so it would appear that these charges do not commute, not even for $c=-2$,
This is in contradiction with the statement that all the KdV and Boussinesq charges are found as bilinears in the symplectic fermion and all mutually commute.

We show in appendix  \ref{app: w0 lambda 0} that the right hand side of \eqref{eq:W0L0} should be set to zero, and the vanishing of the corresponding field puts a constraint on the representations of the $W_3$ algebra at $c=-2$ which is precisely satisfied by those that can be constructed in the symplectic fermion model.

\subsection{Relation to other hierarchies}

The central charge $c=-2$ plays a special role for all the $W_n$ algebras.
These algebras each have minimal models where the central charge is given in terms of two co-prime integers $p$ and $q$
\cite{Fateev:1987zh}:
\begin{align}
    c_{n}(p,q)
    = (n-1)\left(1 - n(n+1)\frac{(p-q)^2}{pq}
    \right)\;,
    \;\;\;\; p,q \geq n
    \;.
\end{align}
In each case we see that $c=-2$ appears as the non-minimal value
\begin{align}
    c_{n}(n-1,n) = -2
    \;.
\end{align}
We have already seen that the symplectic fermion realises the KdV hierarchy related to the Virasoro (or $W_2$ algebra) and the Boussinesq hierarchy related to the $W_3$ algebra. The same does not seem to be true for the higher $W_n$ algebras, at least not straightforwardly, since these all have a non-trivial primary field of weight 4 which is easily seen to be absent from the symplectic fermion model. Some of the structure constants involving the weight 4 field have been worked out in general 
\cite{Gaberdiel:2012ku,Hornfeck:1993kp}, and the weight 4 field decouples from the OPE algebra of the weight 3 field at $c=-2$: in the notation of \cite{Hornfeck:1993kp}, $C_{33}^4$ and $C_{34}^3$ both vanish at $c=-2$, and so it seems likely that symplectic fermion model, if it is a realisation of the $W_n$ algebras at all, is one in which all fields with spin greater than 3 are set to zero.

That this is the case for the  $W_4$ algebra is clear since all the  structure constants were found in \cite{Blumenhagen:1990jv,Kausch:1990bn}, and for the $W_5$ algebra in \cite{Hornfeck:1992tm}. In the first two papers it was stated that the $W_4$ algebra is not defined at $c=-2$ owing to the divergence of the $C_{44}^{(33)}$ structure constant. An alternative viewpoint is that in order for the structure constants of the algebra to be non-singular at $c=-2$, the spin 4 field must be rescaled by a factor of $\sqrt{c+2}$,  and hence itself becomes null at $c=-2$.
With this choice of normalisation, $c=-2$ is a perfectly regular value for the $W_4$ algebra, just one for which the rescaled weight 4 field has zero norm and can be set to zero.

This is the same mechanism for which $c=-22/5$ is a perfectly regular value of the central charge for the $W_3$ algebra, so long as the field $W(z)$ is rescaled, and in which the rescaled weight 3 field has zero norm and can be set to zero.

As a consequence, we believe that the symplectic fermion provides realisations of the integrable hierarchies of the $W_n$ algebras for $n>3$, but in which all the fields of spins greater than 3 are set to zero.

\section{The GGE as a line defect}
\label{sec:defect}

As with the free fermion and Lee-Yang models, the GGE in the symplectic fermion can also be described in the crossed-channel 
(in which the Hamiltonian propagates along the defect - see e.g. \cite{Downing:2023lnp})
as a purely transmitting defect for the symplectic fermion fields.
This can be seen directly form the exact modular transform \eqref{eq: Final Transformation}, \eqref{eq:roots}. In the case $\lambda=1/2$, the trace without the insertion of $(-1)^F$, the transformed partition function can be seen as derived from a set of free fermionic modes with energies determined by a twisted quantisation condition.

We can identify the individual terms in the product, in the case of a right torus, $\hat\tau = i L/R$, as coming from a fermion with energy $E^\pm_n$ propagated along the length $L$,
\begin{align}
    (1 + e^{2\pi i \omega^\pm_l(n)}) = (1 + e^{-L E^\pm_n})\;.
\end{align}
This gives
\begin{align}
    E^\pm_n = - \frac{2\pi}{R}
    {\hat\tau \omega^\pm(n)}
    \;.
\end{align}
If the fermion has the standard dispersion relation, 
then the energies are quantised according to 
\begin{align}
    iPR + i\delta_{\pm}(PR) = 2\pi n i \;,
\end{align}
which is the quantisation condition for a free fermion moving on a circle of length $R$ in the presence of a purely transmitting defect which imparts a phase factor
\begin{align}    
i\delta_\pm(x) = 
\pm (2\pi)\sum_{j=2}^N (\pm1)^j \alpha_j
 \left(\frac{x}{2\pi}\right)^j
\;.
\end{align}
It is natural to associate the transmission factor $\exp(i\delta_\pm)$ with the fields $\chi^\pm$, but simply from the partition function we cannot decide which phase factor is associated to which fermion.

\section{Outlook 
}
In this work we treated the “other easy case” of constructing the KdV hierarchy and deriving the modular transformation of the associated GGEs, emphasising the necessity of enlarging the set of integrable charges to include the bilinear hierarchy. The first “easy” case was the free fermion studied in \cite{Downing:2021mfw}, and evidence for the Symplectic Fermion case was given in \cite{Downing:2024nfb}.

The $W_{1+\infty}$ algebra at $c=1$ admits a realization in terms of bilinear $bc$ ghost fields \cite{Awata:1993vn}, and the central charge can be shifted by deforming the stress tensor with the $U(1)$ generator without changing the $W_{1+\infty}$ algebra itself. Hence the Symplectic Fermion GGE furnishes a $c=-2$ realization of the $W_{1+\infty}$ characters; constructions at negative integral central charge are discussed in \cite{Kac:1995sk}. Dijkgraaf studied the asymptotic modular behaviour of these characters by exploiting a pre-Lie algebra structure that yields a master equation for the modularly transformed characters~\cite{Dijkgraaf:1996iy}.
By contrast, equation~\eqref{eq: Final Transformation} gives an exact expression for the modular $S$-transformation of these characters.

The result \eqref{eq: Final Transformation} is proved rigorously here, but only for the half-twisted and untwisted Symplectic Fermion representations; we have not included the $U(1)$ charge required to isolate the characters that form the modular-invariant bulk state spaces of the Singlet Model. A natural next step is to determine the transformation properties of the generic twisted sectors, with particular attention to the $U(1)$-neutral sectors of the GGE and relations to the analyses in \cite{Gaberdiel:2007jv,Mulokwe:2024zjc}.

While the conjecture in \cite{Downing:2025huv} for the asymptotic behaviour of the GGE was limited to the case of the $W_3$ algebra, it has now been proven in full generality \cite{proof}, removing the constraint from Dijkgraaf's analysis in \cite{Dijkgraaf:1996iy} of the existence of a pre-Lie algebra. It still remains a formula for the asymptotic behaviour though, not the exact transform as we have found here. It would obviously be good to try to extend these asymptotic results to exact results.

It would also be worthwhile to extend this programme to other logarithmic theories, such as the symplectic boson or fractional-level WZW models. An argument analogous to Section \ref{sec: Exact Transformation of Symplectic Fermion GGE} may be unavailable there, but a thermodynamic Bethe ansatz approach (\cite{Downing:2024nfb}) could serve as an alternative route.

Based on the analysis of Section \ref{subsec: Realisation of Hierarchies}, it should further be possible to demonstrate that at $c=-2$ one can directly show the integrability of the quantum Boussinesq hierarchy in the sense of \cite{DiFrancesco:1991xm}. 

Another concrete direction is to construct explicitly the spectrum and Hamiltonian of the defect Hilbert space associated with the modular transformation, following the methods of \cite{Downing:2023lnp}.

It would also be compelling to use our transformation formula to derive a Cardy-like formula for the GGE with $W_0$ inserted and compare it to the $\mathcal{W}_3$ black hole entropy of \cite{Gutperle:2011kf}. Our results provide a possibility of calculating it at finite $c$.

It would be useful to understand how the Symplectic Fermion GGE is realised on the lattice. In \cite{Nigro:2009si} the integrals of motion are understood directly on the lattice (Critical Dense Polymers). Earlier we interpreted the GGE as a translation-invariant line defect, motivated by the free-fermion GGE as a defect \cite{Downing:2023lnp}. A lattice interpretation of this defect, or an interpretation in terms of the $U_q(\mathfrak{sl}(2))$ spin chain at $q=i$, would therefore be desirable (see e.g. \cite{Chernyak:2022wic,Baseilhac:2025upc}).

In \cite{Downing:2023uuc} the thermal correlators of bilinear tensor currents in massive free fermions were shown, in the conformal limit, to correspond to the KdV hierarchy of the $c=\tfrac{1}{2}$ free-fermion CFT, and to be described by massive Maass--Jacobi forms. It should be possible to analyse massive symplectic fermions similarly and to obtain analogous modular-type results for currents associated not only with KdV but also with the full Bilinear hierarchy.

Recent work \cite{Brizio:2025fms} studies GGEs with fractional-spin charges arising in graded integrable field theories with an internal $\mathbb{Z}_n$ structure. It would be interesting to explore this framework for the Triplet model/Symplectic Fermions, in close analogy with the graded Ising/free-fermion example in that paper. For the graded Lee--Yang model the fractional GGE admits a TBA description derivable from a deformed cubic oscillator via the ODE/IM correspondence \cite{Masoero:2010is}; the authors of \cite{Brizio:2025fms} suggest extending this correspondence to other graded theories. If a graded Triplet model exists, one might therefore expect a corresponding ODE description.

It is natural to ask whether the ODE/IM correspondence can be adapted to GGEs involving KdV charges. The KdV GGE transformation is now known exactly at $c=\tfrac{1}{2}$ and $c=-2$ (both free theories), so a deformed ODE/IM approach may be tractable in these cases. Both central charges are associated with ODEs having a quadratic potential \cite{Dorey:2007ti}, suggesting a unified treatment. Indeed, $c=-2$ was noted as an ``easy'' case in \cite{Bazhanov:1998wj} (the ``free-fermion point'').

The Symplectic Fermion realisation of the $W_3$ algebra permits the construction of GGEs tied to $W$-algebra hierarchies. Since the ODE/IM correspondence for $W_N$ algebras is well developed (see e.g. \cite{Dorey:1999pv,Dorey:2006an}) and allows extraction of integrals-of-motion eigenvalues \cite{Kudrna:2025bzg}, it is natural to seek an extension of ODE/IM to GGEs based on $W$-algebras. The polynomial relations \eqref{eq:roots} (and their analogues in the $c=\tfrac{1}{2}$ free-fermion case \cite{Downing:2021mfw}) admit a TBA interpretation; combined with discussions of reversing the ODE/IM map \cite{Fioravanti:2021bzq}, this could lead to a precise correspondence for both KdV GGEs and $W$-algebra GGEs.

\acknowledgments
We would like to thank Sujay Ashok, Max Downing, Tanmoy Sengupta and Adarsh Sudhakar for very many discussions on GGEs and modular properties and comments on the manuscript.
We would both like to thank Azat Gainutdinov for extensive discussions on the symplectic fermion and for pointing out the connection to lattice models.
In addition, FK would like to thank Joseph Smith, and GW would like to thank Ingo Runkel, for further discussions.

FK is supported by an STFC studentship
under grant ST/Y509279/1.

GW would like to thank the Isaac Newton Institute for Mathematical Sciences, Cambridge, for support and hospitality during the programme Quantum field theory with boundaries, impurities, and defects, where work on this paper was undertaken. This work was supported by EPSRC grant EP/Z000580/1.
GW also thanks the Fachbereich Mathematik, Hamburg University for hospitality where work on this paper 
was also undertaken, and the Deutsche Forschungsgemeinschaft (DFG, German Research Foundation)
under Germany’s Excellence Strategy - EXC 2121 “Quantum Universe” - 390833306, and the
Collaborative Research Center - SFB 1624 “Higher structures, moduli spaces and integrability” - 506632645, for support. GW was, in addition, supported by the STFC under grant
ST/T000759/1.

For the purposes of open access, the authors have applied a Creative Commons Attribution (CC
BY) licence to any Accepted Author Manuscript version arising from this submission.

\appendix

\section{(Quasi-) Modular Forms}\label{app: Modular Forms}
In this appendix we will list the relevant facts about modular forms that appear in this paper. Proofs of the following statements can be found in \cite{Zagier2008} and most of the notation will be the same. 

The full modular group will be denoted by
\begin{equation}
    \Gamma_1 = \slz.
\end{equation}
We define the congruence subgroup $\Gamma(N)$ by
\begin{equation}
    \Gamma(N):= \{\g \in \G_1 |\quad \gamma \equiv \bb{I}_2 ,\mod N\}.
\end{equation}
Here $\bb{I}_2$ is the $2\times2$ identity matrix.

A holomorphic function $f(\t)$ define on the upper half plane is said to be a modular form of weight $k$ for the group $\Gamma$, which for us is $\Gamma_1$ or $\Gamma(2)$, if it has the transformation property
\begin{equation}
    f\left(\frac{a\t+b}{c\t+d}\right) = (c\t+d)^k f(\t),\quad \begin{pmatrix}
        a&b\\c&d
    \end{pmatrix} \in \Gamma.
\end{equation}
The space of weight $k$ modular forms for $\Gamma$ we call $M_k(\Gamma)$. The group $\Gamma_1$ is generated by the matrices
\be\label{eq: Representative for S}
     T = \begin{pmatrix}1&1\\0&1\end{pmatrix} \;,\quad 
     S = \begin{pmatrix} 0&-1\\1&0 \end{pmatrix}\;,
\ee 
hence we only need to check that a function transforms as a modular form under
\be
    T:\tau\mapsto\tau+1 \;,\quad S:\tau\mapsto\frac{-1}{\tau}\;,
\ee
to verify it is an element of $M_k(\Gamma_1)$. The group $\Gamma(2)$ is 
generated by the matrices
\be
     \begin{pmatrix}1&2\\0&1\end{pmatrix} \;,\quad \begin{pmatrix} 1&0\\2&1 \end{pmatrix}\;,\;\;-\bb{I}_2,
\ee 
hence we only need to check that a function transforms as a modular form under
\be
    \tilde{T}:\tau\mapsto\tau+2 \;,\quad \tilde{S}:\tau\mapsto\frac{\t}{2\t+1}\;,
\ee
to verify that is belongs to $M_{k}(\Gamma(2))$. The space $M_{2k}(\Gamma_1)$ is generated by the Eisenstein series (see below) and
the space $M_{2k}(\Gamma(2))$ is generated by the Jacobi theta functions (see \cite{Zagier2008} for definitions of the
Jacobi theta functions).

The Eisenstein series $E_{2k}(\t)$ are elements of $M_k(\Gamma_1)$, $k =2,3,...$, and they are defined by
\begin{equation}
    E_{2k}(\tau)=1+\frac{2}{\zeta(1-2k)}\sum_{n=0}^\infty\frac{n^{2k-1}q^n}{1-q^n}\;,\;\;\;q=e^{2\pi i\tau}\;.
\end{equation}
The $k=1$ Eisenstein series', $E_2(\t)$ is quasi-modular. It has the following modular transformation property under $\Gamma_1$
\begin{equation}
E_{2}\left(\frac{a\tau+b}{c\tau+d}\right)=(c\tau+d)^2E_2(\tau)-\frac{6i}{\pi}c(c\tau+d)\;.
\end{equation}
It is worth noting that, given $\tilde{T} = T^2$ and $\tilde{S} = (TST)^2$, one can show that
\begin{equation}
    E_{2k}(2\t),\; E_{2k}(\t), \; E_{2k}(\tfrac{\t}{2}) \in M_{2k}(\Gamma(2)).
\end{equation}
We use $E_2$ to define higher quasi-modular forms by the following. The space of quasi-modular forms of weight $k$ and depth $p$ for a group $\Gamma$, $\tilde{M}^{(p)}_k(\Gamma)$ is define to be
\begin{equation}
    \tilde{M}^{(p)}_k(\Gamma) = \bigoplus_{r=0}^{p}M_{k-2r}(\Gamma)\cdot E_2^r.
\end{equation}
The above is understood in such a way that any element of that set should have a coefficient of $E_2^p$ non-zero. Finally we define the Serre derivative. The Serre derivative acting on a modular form $F_k(\tau)$ of weight $k$ is defined to be
\be
    D_k F_k(\tau)=\frac{1}{2\pi i}\frac{d}{d\tau}F_k(\tau)-\frac{k}{12}E_2(\tau)F_k(\tau)\;.
\ee
By using the transformation of $\frac{1}{2\pi i}\frac{d}{d\tau} =  q \frac{d}{dq}$ under a modular transform we can see that $D_kF_k(\tau)$ is a modular form of weight $k+2$. From this it is not difficult to see that
\begin{equation}
    D_0F_k \in \tilde{M}_{k+2}^{(1)}(\Gamma).
\end{equation}

\section{Properties of Roots}\label{app: Properties of Roots}
The roots $\om^{\pm}(\kappa)$ satisfy the equation
\begin{equation}\label{eq: root equationa appendix}
\om = \om_l^{\pm}(\kappa): \quad\sum_{n=1}^N(\pm1)^{n+1}(\t^n\a_n\om^n)=\kappa.
\end{equation}
We're assuming that we're working in the restricted domain \eqref{eq: Restricted Domain}, which we repeat here. Let $\Q_{2K-1}$ be the highest odd-spin charge appearing the GGE \eqref{eq: GGE alphas.}, and let there be only finitely many even spin charges appearing in the GGE $\{\Q_{2n_i}\}$ with spin greater than $2K-1$. Then the restricted domain is the one which has constrained chemical potentials for all of the chemical potential and not just the highest ones.
\begin{equation}\label{eq: Restricted domain appendix}
\begin{split}
&\Im(\a_{2n-1}\t^{2n-1})\geq0, \quad \forall\,\,\, n<K,\\
&\Im(\a_{2K-1}\t^{2K-1})>0,  \\
&\Im(\a_{2n}\t^{2n})=0, \quad \forall \,\,\,2n<N.\\
\end{split}
\end{equation}
It is easy to show that if $\om_l^\pm(\kappa)$ solves the above equations, then so does $-\om_l^{\mp}(-\kappa)$, that is we can pair up roots so that
\begin{equation}
    \om_l^\pm(\kappa) = - \om_l^\mp(-\kappa).
\end{equation}
We call by $\om_0^{\pm}(\kappa)$ the distinguished root continuously connected to $0$ when $\kappa =0$.

Let us show that for all $l$, and all $\kappa >0$, that the roots never hit the real axis. We can show this by contradiction, if any roots were to hit the real axis, then for some $\kappa>0$ we would have $\Im(\om_l^\pm(\kappa))=0$. Taking the imaginary part of \eqref{eq: root equationa appendix}, we see that
\begin{equation}
    \sum_{n=1}^N(\pm1)^{n+1}\Im(\t^n \a_n) (\om_l^\pm(\kappa))^n = 0,
\end{equation}
and since we work in the domain \eqref{eq: Restricted domain appendix}, we require $z_j^\pm(\om)=0$, which can only happen if $\omega=0$, a contradiction.

Let us show that for $l=0$, and all $\kappa>0$, that the root $\om_0^\pm(\kappa)$ always has negative imaginary part. Assuming $\om_0^\pm(\kappa)$ has a series expansion in $\kappa$, we can show by substitution into \eqref{eq: root equationa appendix} that the leading piece is $\om_0^\pm(\kappa) = \frac{\kappa}{\a_1 \t}+O(\kappa^2)$. Therefore, the domain \eqref{eq: Restricted domain appendix} tells us that for small enough $\kappa$, $\om_0^\pm(\kappa)$ will have a negative imaginary part, and since the roots never cross the real axis, $\om_0^\pm(\kappa)$ will always have negative imaginary part.

Let us show that the roots $\om_l^\pm(\kappa)$ will have opposite sign imaginary parts for $\kappa >0$ and $l>0$. This simply follows from that fact that we know that for $\kappa>0$, the roots never hit the real axis, and that $\om_l^\pm(0) = -\om_l^\mp(0)$. Therefore if $\Im(\om^\pm_l(\kappa))>0$, then $\Im(\om^\mp_l(\kappa))<0$.

Let us find the asymptotic behaviour of the roots as $\kappa \to \infty$. We can assume that the leading behaviour as $\kappa$ becomes large is some power law
\begin{equation}
    \om_l^\pm(\kappa) = A^\pm \kappa^{\nu_{\pm}}.
\end{equation}
Substituting in to \eqref{eq: Restricted domain appendix} gives us the leading order behaviour
\begin{equation}
    (\pm1)^{N+1}\a_N (\t A^\pm \kappa^{\nu_\pm})^N \sim \kappa.
\end{equation}
Hence
\begin{equation}
\begin{split}
    \nu_\pm&= \frac{1}{N}, \quad
    A^+ = \frac{e^{\frac{2\pi i k}{N}}}{(\a_N\t^N)^{\frac{1}{N}}},\quad
    A^- = \frac{e^{\frac{(i \pi+2\pi i k)(N+1)}{N}}}{(\a_N\t^N)^{\frac{1}{N}}},\\
    k&=1,...,N-1.
\end{split}
\end{equation}
The important thing in our calculation is that the indeed the roots do grow to infinity, and as shown previously, maintain the sign of the imaginary parts as $\kappa>0$ grows.

\section{\boldmath $[W_0,\Lambda_0]$ at $c=-2$}\label{app: w0 lambda 0}

The fact that both the KdV and Boussinesq hierarchies appear as subsets of the same bilinear hierarchy at $c=-2$ suggests an apparent contradiction, namely
\begin{equation}
    [W_0,\Lambda_0]=0,
\end{equation}
which is not generically true. We will show that this relation indeed holds only at $c=-2$, and only in representations connected to the Symplectic Fermion (and hence to related orbifold models \cite{Creutzig:2013hma}).

It is well known that the commutator of two zero modes can be expressed as the zero mode of a field. Let $A(z)$ and $B(w)$ be fields of weight $h_A$ and $h_B$, then by rewriting the planar zero-modes as contour integrals of the fields, and looking at the OPE between $A$ and $B$, it is not difficult to show that
\begin{equation}
    [A_0,B_0] = X_0,
\end{equation}
where $X(z)$ is the field associated to the state
\begin{equation}
    \ket{X} = \sum_{p=0}^{h_A-1} 
    (-1)^p
    \frac{(1-h_A)_p}{ p!\,(2-h_A-h_B)_p}
    L_{-1}^p A_{p-h_A + 1}\ket{B}.
\end{equation}
Setting $A = W$, and $B = \Lambda = (TT)-\frac{3}{10}\del^2T$, we find that
\begin{equation}
    \ket{X} = -\frac{12}{5}\left(L_{-3}W_{-3}-\tfrac{2}{3} W_{-4}L_{-2}+\tfrac{1}{2} W_{-6}\right)\ket{0}.
\end{equation}
The crucial observation is that at $c=-2$, this state becomes null, and so $[W_0,\Lambda_0]=0$. By null we mean that the state satisfies the same conditions of a highest weight state that generates a highest weight representation of the $W_3$ algebra.
\begin{equation}
    W_{n>0}\ket{X} = L_{n>0}\ket{X} =0, \quad L_0\ket{X} = h_{X}\ket{X},\quad W_0\ket{X} = w_{X}\ket{X}
\end{equation}
One can show that there exists another null vector at level $6$. That is
\begin{equation}
    \left(\tfrac{7}{4} L_{-4}L_{-2}+\tfrac{19}{32} L_{-3}^2+L_{-2}^3-\tfrac{9}{8} W_{-3}^2-2 L_{-6}\right)\ket{0}.
\end{equation}
Acting with the planar zero-mode of the associated field on a generic highest weight state with weights $h$ and $w$ and setting the result to zero imposes the following constraint
\begin{equation}
    h^2 + 8h^3 - 9w^2=0.
\end{equation}
This is confirmation of an earlier study of $W_3$ representations \cite{Wang:1997ndk} at $c=-2$. For a generic twisted representation $\bb{L}_{\lambda}$ of the Symplectic Fermion, the conformal weight is
\begin{equation}
    h(\lambda) = -\tfrac{1}{2}\lambda(1-\lambda),
\end{equation}
and from \eqref{eq: Regularisation Constant} the $W_0$ eigenvalue is
\begin{equation}
    w(\lambda)=-\tfrac{1}{6}\lambda(\lambda-1)(2\lambda-1).
\end{equation}
It is straightforward to check that these satisfy the above constraint, so indeed all Symplectic Fermion representations of the $W_3$ algebra exhibit $[W_0,\Lambda_0]=0$.

From this perspective, it becomes natural to ask whether it is “obvious’’ within the $W_3$ algebra that all KdV charges commute with $W_0$. One definition of the KdV charges is via perturbation of the CFT \cite{Mathieu:1996un} by the primary field with Kac labels $(1,3)$, denoted $\phi_{(1,3)}$, so that
\begin{equation}
    [\I^{\text{KdV}}_{2n-1},\int \phi_{(1,3)}\,d^2z] = 0.
\end{equation}
This field has conformal weight
\begin{equation}
    h_{(1,3)} = 2t - 1,
\end{equation}
where $t$ is defined by
\begin{equation}
    c = 13 - 6\left(t+\tfrac{1}{t}\right).
\end{equation}
At $c=-2$, this yields
\begin{equation}
    h_{(1,3)} = 0 \quad \text{or} \quad 3.
\end{equation}
For generic central charge the two choices are equivalent in that they provide the same set of conserved densities. However, in our case the weight-$3$ primary is precisely the field $W(z)$. Thus all KdV charges commute with the mode $W_0$.

\section{Pre-Lie Algebra}\label{subsec: Pre-Lie}
In this section we will show that the bilinear fields \eqref{eq: Bilinear Fields} at $c=-2$ satisfy the same pre-lie algebra relations that were found in \cite{Dijkgraaf:1996iy} for the $c=1$ two-component fermion. This result is not surprising, but it is still important to show since in the companion paper \cite{Downing:2025huv} we rely on these results to prove that main conjecture of that paper at $c=-2$.

Take any two chiral fields $A(z)$ and $B(z)$. Their OPE is then
\begin{equation}
    A(z)B(w) \sim \sum_{k>0}\frac{(AB)_k}{(z-w)^k}.
\end{equation}
For the pre-lie algebra structure, one is only interested in the first and second order poles in the OPE. Consider the following OPE
\begin{equation}
(\del^{n-2-k}\chi^-\del^k \chi^+)(z)(\del^{m-2-l}\chi^-\del^{l}\chi^+)(w) \sim ...+ \frac{\CX^{m;l}_{n;k}(w)}{(z-w)} +\frac{\CY^{m;l}_{n;k}(w)}{(z-w)^2} ,
\end{equation}
where the dots denote more singular terms. It is not difficult to calculate that
\begin{equation}
\begin{split}
\CX_{n;k}^{m;l}(w)&=(-1)^k \left((\del^{n-l+m-3}\chi^-\del^l \chi^+)(w) - (-1)^n(\del^{l + n-1}\chi^+\del^{m-2-l}\chi^-)(w)\right),\\
\CY_{n;k}^{m;l}(w)&=(-1)^k \biggl((k+m-l-1)(\del^{n-l+m-4}\chi^-\del^l \chi^+)(w) \\
&\qquad- (-1)^n(n-k+l-1)(\del^{l + n-2}\chi^+\del^{m-2-l}\chi^-)(w)\biggr).
\end{split}
\end{equation}
Recall the form of the Bilinear fields \eqref{eq: Bilinear Fields}:
\begin{equation}
\begin{split}
B_n(z) &=  \sum_{k=0}^{n-2} (-1)^k\Omega_{n;k}(\del^{n-2-k}\chi^- \del ^k \chi^+)(z),\\
    & = (\del^{n-2}\chi^- \chi^+)(z)  \text{ mod }\del,
\end{split}
\end{equation}
where
\begin{equation}
    \Omega_{n;k}=\left[\frac{1}{(n-2)!}\begin{pmatrix}
        2n-2\\n-1
    \end{pmatrix}\right]^{-1}  \frac{ \binom{n}{k+1}}{(n-2-k)!k!}.
\end{equation}
Then we can use the expressions for $\CX_{n;k}^{m;l}$ and $\CX_{n;k}^{m;l}$ to fix the first and second order poles in the expansions between the bilinear fields. Indeed we find
\begin{equation}
    (B_nB_m)_1 = \sum_{k=0}^{n-2}\sum_{l=0}^{m-2}(-1)^{k+l}\Omega_{n;k}\Omega_{m;l}\CX^{m;l}_{n;k},
\end{equation}
\begin{equation}
    (B_nB_m)_2 = \sum_{k=0}^{n-2}\sum_{l=0}^{m-2}(-1)^{k+l}\Omega_{n;k}\Omega_{m;l}\CY^{m;l}_{n;k}.
\end{equation}
Using the identity\footnote{We need to apply this formula twice in the case of the first order pole}:
\begin{equation}
    (\del^a\chi^-\del^b \chi^+)(z) = \sum_{c=0}^{b-1}\del\biggl((-1)^c(\del^{a+c}\chi^-\del^{b-c-1}\chi^+)(z)\biggr) + (-1)^b(\del^{a+b}\chi^-\,\, \chi^+)(z),
\end{equation}
and also the fact that:
\begin{equation}
    \sum_{k=0}^{n-2}\Omega_{n;k}=1,
\end{equation}
one can show that:
\begin{equation}
    (B_nB_m)_2 = (m+n-2) B_{n+m-2} \text{ mod }\del.
\end{equation}
and also:
\begin{equation}
(B_nB_m)_1 = \del \biggl((n-1)B_{m+n-2} + \del(...) \biggr) = 0 \mod \del
\end{equation}
We see precisely the same pre-lie algebra structure found in \cite{Dijkgraaf:1996iy}. We construct a connection in the space of chiral fields modulo total derivatives
\begin{equation}
    \nabla_{B_m}B_n  = \del^{-1}((B_nB_m)_1) = (n-1)B_{m+n-2} \mod \del
\end{equation}
which defines then the pre-lie bracket
\begin{equation}
    [B_m , B_n]_2 := \nabla_{B_m}B_n - \nabla_{B_n}B_m = (n-m)B_{m+n-2} \mod \del 
\end{equation}
In the companion paper \cite{Downing:2025huv}, we use the second order pole to conjecture a general form for the asymptotic modular transformation of the GGE.





\bibliographystyle{JHEP}
\bibliography{biblio.bib}
\end{document}